\newif\ifconfver

\newif\ifplainver  
\plainvertrue

\newif\ifhide  
\hidetrue

\ifplainver
    \confverfalse   
\fi

\ifconfver
     \documentclass[10pt,twocolumn,twoside]{IEEEtran}
\else
    \ifplainver
        \documentclass[11pt]{article}
        \usepackage{fullpage}
    \else
        \documentclass[12pt,draftcls,onecolumn]{IEEEtran}
    \fi
\fi

\usepackage{calc,amsfonts,amssymb,amsmath,bm,url,color,theorem,graphicx,cite}
\usepackage{psfrag,float}
\usepackage{algorithm}
\usepackage{algorithmic}
\usepackage{soul}
\usepackage{enumerate}
\usepackage{bbm}
\usepackage{multirow}
\usepackage{array}
\usepackage{epstopdf}
\usepackage{diagbox}
\usepackage{cases}
\usepackage{makecell}
\usepackage{subfigure}

\newlength{\articlesectionshift}%
\let\LaTeXStandardSection\section

\makeatletter

\def\@IEEEdestroythesectionargument#1{\LaTeXStandardSection{#1}}%

\@addtoreset{section}{titlecounter}

\newlength{\customfigwidth}
\newcolumntype{M}[1]{>{\centering\arraybackslash}m{#1}}

\definecolor{orange}{RGB}{255,107,0}

\definecolor{cpink}{rgb}{0.7, 0.11, 0.11}

\newtheorem{Fact}{Fact}

\newtheorem{Asm}{Assumption}

\newcommand\bx{\ensuremath{{\bm x}}}
\newcommand\by{\ensuremath{{\bm y}}}

\newcommand\bh{\ensuremath{{\bm h}}}
\newcommand\bH{\ensuremath{{\bm H}}}

\newcommand\bz{\ensuremath{{\bm z}}}

\newcommand\br{\ensuremath{{\bm r}}}

\newcommand\bX{\ensuremath{{\bm X}}}
\newcommand\bZ{\ensuremath{{\bm Z}}}

\newcommand\ba{\ensuremath{{\bm a}}}

\newcommand\bF{\ensuremath{{\bm F}}}

\newcommand\bu{\ensuremath{{\bm u}}}

\newcommand\bs{\ensuremath{{\bm s}}}

\newcommand{\Rbb}{\mathbb{R}}

\newcommand{\Cbb}{\mathbb{C}}

\newcommand{\setX}{\mathcal{X}}

\newcommand{\setS}{\mathcal{S}}

\newcommand{\Exp}{\mathbb{E}}

\newcommand\jj{\ensuremath{{\mathfrak j}}}

\newcommand{\bzero}{{\bm 0}}

\newcommand{\Tsf}{\top}
\newcommand{\Hsf}{\mathsf{H}}

\newcommand{\beq}{\begin{equation}}
\newcommand{\eeq}{\end{equation}}

\begin{document}

\bibliographystyle{IEEEtran}

\newcommand{\papertitle}{
A Spatial Sigma-Delta Approach to Mitigation of Power Amplifier Distortions in Massive MIMO Downlink

}

\newcommand{\paperabstract}{
In massive multiple-input multiple-output (MIMO) downlink systems,
the physical implementation of the base stations (BSs) requires the use of cheap and power-efficient power amplifiers (PAs) to avoid high hardware cost and high power consumption.
However, such PAs usually have limited linear amplification ranges.
Nonlinear distortions arising from operation beyond the linear amplification ranges can significantly degrade system performance.
Existing approaches to handle the nonlinear distortions, such as digital predistortion (DPD), typically require accurate knowledge, or acquisition, of the PA transfer function.
In this paper, we present a new concept for mitigation of the PA distortions.
Assuming a uniform linear array (ULA) at the BS,
the idea is to apply a Sigma-Delta ($\Sigma \Delta$) modulator to spatially shape
the PA distortions to the high-angle region.
By having the system operating in the low-angle region,
the received signals are less affected by the PA distortions.
To demonstrate the potential of this spatial $\Sigma \Delta$ approach, we study
the application of our approach to the multi-user MIMO-orthogonal frequency division modulation (OFDM)  downlink scenario.
A symbol-level precoding (SLP) scheme and a zero-forcing (ZF) precoding scheme, with the new design requirement by the spatial $\Sigma \Delta$ approach being taken into account, are developed.
Numerical simulations are performed to show the effectiveness of the developed $\Sigma \Delta$ precoding schemes.
}


\ifplainver


    \title{\papertitle}

    \author{Yatao Liu$^\dag$,  Mingjie Shao$^{\dag\S}$ and Wing-Kin Ma$^\dag$ \\~\\
    $^\dag$Department of Electronic Engineering, The Chinese University of Hong Kong, \\
    Hong Kong SAR, China\\
    $^\S$School of Information Science and Engineering,
Shandong University, Qingdao, China\\~\\
E-mails: ytliu@ee.cuhk.edu.hk, mingjieshao@sdu.edu.cn, wkma@cuhk.edu.hk
\thanks{This work was supported in part by a General Research Fund (GRF) of Hong
Kong Research Grant Council (RGC) under  Project ID CUHK 14208819.}
    }

    \maketitle

    \begin{abstract}
    \paperabstract
    \end{abstract}

\else
    \title{\papertitle}

    \ifconfver \else {\linespread{1.1} \rm \fi

    \author{XXX, XXX, AND XXX
    }

    \maketitle

    \ifconfver \else
        \begin{center} \vspace*{-2\baselineskip}
        \end{center}
    \fi

    \begin{abstract}
    \paperabstract
    \end{abstract}


    \begin{IEEEkeywords}\vspace{-0.0cm}
       Massive MIMO, MIMO-OFDM, nonlinear distortion, power amplifier, Sigma-Delta modulation, symbol-level precoding.
    \end{IEEEkeywords}

    \ifconfver \else \IEEEpeerreviewmaketitle} \fi

 \fi

\ifconfver \else
    \ifplainver \else
        \newpage
\fi \fi

\maketitle

\section{Introduction}

Over the past decade, massive multiple-input multiple-output (MIMO) systems, which employ large antenna arrays at the base station (BS), have attracted tremendous interest from both academia and industry, due to its potential of offering orders of magnitude improvements in spectral and power efficiencies compared to conventional MIMO systems.
Despite such potential, the practical implementation of massive MIMO requires the use of low-cost and power-efficient power amplifiers (PAs) at the BS;
we should bear in mind that the number of PAs, and the subsequent hardware cost and power consumption, scale with the antenna array size.
Such PAs generally exhibit nonlinear transfer characteristics, with limited linear amplification ranges; and operation beyond the linear amplification ranges can result in nonlinear distortions to the transmitted signals~\cite{cripps2006rf}.
Researchers have investigated the impact of the PA distortions
on the massive MIMO downlink,
and it has been shown that the PA distortions can lead to substantial performance degradation~\cite{mollen2016waveforms,larsson2018out,anttila2019antenna,moghadam2018energy,moazzen2019performance}.


As a result, mitigating the PA distortion effects has become an important research problem.
Digital predistortion (DPD) is a promising concept to deal with PA nonlinearity~\cite{ghannouchi2009behavioral,guan2014green,katz2016evolution}. It applies an inverse response, called a predistorter, at the input of the PA to equalize the nonlinear PA transfer characteristics.
In order to do that, DPD requires acquisition of the nonlinear PA transfer function.
There is a certain hardware implementation cost for DPD with each PA~\cite{wood2017system}.
While DPD has been used in small- or medium-scale MIMO, it can be expensive to use in massive MIMO.
As a compromise, one can impose certain power constraints on the PA input signals, such as peak power constraints or
power back-off~\cite{schenk2008rf,jedda2017precoding,spano2017symbol},
peak-to-average power ratio (PAPR) constraints~\cite{studer2013aware,bao2016efficient,yao2018semidefinite,qin2021low,spano2017symbol},
and constant-envelope constraints~\cite{mohammed2013per,pan2014constant,shao2019framework,domouchtsidis2020constant}, to keep the PAs operating in a limited linear amplification region.
This approach does not have additional hardware requirement on top of the PAs (unlike DPD),
but it may result in either lower PA power efficiency (e.g., power back-off) or more difficult signal designs (e.g., those with PAPR constraints).
Another approach is to
take the PA distortions into account when designing the transmitted signals.
One can model the PA distortions, together with other hardware impairments (such as phase noise and quantization errors), as additive Gaussian noise, where the noise power scales with the PA input power; see, e.g.,~\cite{bjornson2012optimal,brandt2014weighted,zarei2017multi}.
However,  such Gaussian noise model
can be inaccurate \cite{larsson2018out,anttila2019antenna}.
A more advanced method is to apply the Bussgang theorem to approximate the nonlinear PA model as a linear one---modeling the PA nonlinear effects via signal scalings and uncorrelated additive distortion noise;
see, e.g.,~\cite{aghdam2020distortion,jee2020precoding,jee2021joint}.

In this work,
we present a different concept for combating the PA distortion effects.
The idea is to apply Sigma-Delta ($\Sigma \Delta$) modulation~\cite{aziz1996overview}.
$\Sigma \Delta$ modulation appears most frequently in analog-to-digital or digital-to-analog converters of temporal signals.
It aims at reducing the quantization noise effects on oversampled, low-frequency, temporal signals.
The principle is to apply a specific feedback loop to shape the quantization noise to high frequency.
Consequently, the shaped quantization noise is pushed away from the low-frequency signal in the frequency-domain, and we can remove much of the quantization noise via low-pass filtering.
The reader is referred to the overview paper~\cite{aziz1996overview} and the references therein for details.
More recently, the $\Sigma \Delta$ principle has also been explored in the spatial domain.
Specifically, by utilizing spatial oversampling (i.e., using sub-half-wavelength inter-antenna spacings in a uniform linear antenna array) and a spatial $\Sigma \Delta$ modulator (with feedback loops between adjacent antennas), the quantization noise can be shaped to high spatial frequency (i.e., angle) and the signals in the low-angle region will be less affected by the quantization noise.
This spatial $\Sigma \Delta$ idea has been used for quantization noise reduction in a number of  wireless communication and radar applications, such as
signal detection or channel estimation in the uplink~\cite{corey2016spatial,barac2016spatial,nikoofard2017low,madanayake2017improving,rao2019massive,pirzadeh2020spectral,rao2021massive}, and  beamforming or precoding in the downlink~\cite{scholnik2004spatio,krieger2013dense,shao2019one,shao2020multiuser}.
These works show the effectiveness of spatial $\Sigma\Delta$ modulation in quantization noise mitigation.

As the key contribution of this work, we advocate to use spatial $\Sigma \Delta$ modulation to handle the PA distortion problem in the downlink transmission.
To the best of our knowledge, this is the first work that explores the use of spatial $\Sigma \Delta$ modulation to combat the PA distortions.
Assuming a uniform linear array (ULA) at the BS, we show that the PA distortions can be effectively mitigated by applying a spatial $\Sigma \Delta$ modulator that shapes the PA distortions to the high-angle region.
It should be emphasized that, different from the $\Sigma \Delta$ modulator in~\cite{shao2019one}, which is implemented in the digital domain, the $\Sigma \Delta$ modulator concept presented in this study is implemented in the analog domain.
It is worth noting that our spatial $\Sigma \Delta$ approach does not require precise knowledge, or acquisition, of the PA transfer function.
Our approach is based on the assumption that we know the worst-case magnitude of the PA distortion relative to the ideal linear amplification response.
That worst-case magnitude can be a guessed one in practice.


To demonstrate the spatial $\Sigma \Delta$ modulation concept for PA distortion mitigation, we design suitable precoding schemes under this new concept.
Our scenario of interest is that of MIMO-OFDM, which is considered a more realistic scenario than the standard frequency-flat MIMO scenario.
As we will show, the spatial $\Sigma \Delta$ modulation concept leads to amplitude constraints with the transmitted signals.
We develop a zero-forcing (ZF) scheme and a symbol-level precoding (SLP) scheme for the precoding design, both assuming quadrature amplitude modulation (QAM) constellations.
SLP is an emerging precoding paradigm that optimizes symbol-level performance metrics. We formulate our SLP design as a maximum detection probability (DP) problem subject to signal amplitude constraints, which is a large-scale convex problem. To efficiently solve this problem, we custom build an alternating direction method of multipliers (ADMM) algorithm.

Relevant existing SLP studies should be mentioned.
There have been numerous SLP schemes developed for multi-user MIMO, but without OFDM~\cite{spano2017symbol, mohammed2013per,pan2014constant,shao2019framework, masouros2015exploiting,alodeh2015constructive,alodeh2017symbol,liu2021symbol,jacobsson2017quantized,sohrabi2018one}.
There are also some studies for multi-user MIMO-OFDM~\cite{studer2013aware,bao2016efficient,yao2018semidefinite,qin2021low,askerbeyli20191,domouchtsidis2020constant,jacobsson2018nonlinear,tsinos2020symbol}.
The design criteria of the existing SLP schemes include
total power reduction~\cite{masouros2015exploiting,alodeh2015constructive,alodeh2017symbol,liu2021symbol},
per-antenna power minimization~\cite{spano2017symbol,liu2021symbol},
PAPR reduction~\cite{studer2013aware,bao2016efficient,yao2018semidefinite,qin2021low},
one-bit constraints~\cite{jacobsson2017quantized,sohrabi2018one,askerbeyli20191,shao2019framework},
constant-envelope constraints~\cite{mohammed2013per,pan2014constant,shao2019framework,domouchtsidis2020constant},
and phase quantized constant-envelope constraints~\cite{jacobsson2018nonlinear,tsinos2020symbol,shao2019framework}.
However, we are unaware of any existing SLP scheme that best fits the requirement of limited signal amplitudes introduced by our spatial $\Sigma \Delta$ approach and under QAM constellations.
This calls for the need to custom design an SLP scheme for our spatial $\Sigma \Delta$ approach.
Our numerical results will show that the proposed scheme, combined with the spatial $\Sigma \Delta$ approach, gives promising performance.

Our notations are as follows.
We use lowercase letters (e.g., $x$), boldfaced lowercase letters (e.g., $\bx$), and boldfaced capital letters (e.g., $\bX$) to denote scalars, column vectors, and matrices, respectively;
$\bX^\Tsf$ and $\bX^\Hsf$ represent the transpose and Hermitian transpose of $\bX$, respectively;
$\jj = \sqrt{-1}$ is the imaginary unit;
$\Re(\bx)$ and $\Im(\bx)$ denote the real and imaginary components of $\bx$, respectively;
$\mathbb{R}$ and $\mathbb{C}$ are the sets of all real and complex numbers, respectively;
$\mathcal{U}{[a,b]}$ denotes the uniform distribution on $[a,b]$;
$\Pi_{\setX}(\bx) \in \arg \min_{\by \in \setX} \|\bx-\by\|_2^2$ is a projection of $\bx$ onto the set $\setX$.

\section{Background}

\subsection{PA Model}
\label{sec:PA_mod}


\begin{figure}[t]
\centering
\subfigure[AM-AM conversion.]{
\label{fig:PA_response_am}
\includegraphics[width=0.5\linewidth]{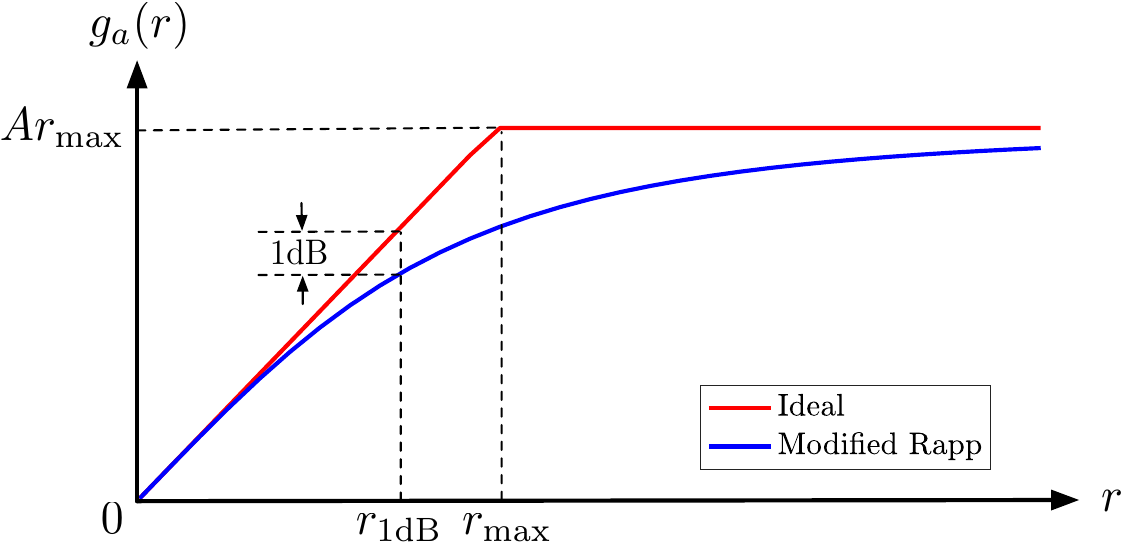}}
\subfigure[AM-PM conversion.]{
\includegraphics[width=0.5\linewidth]{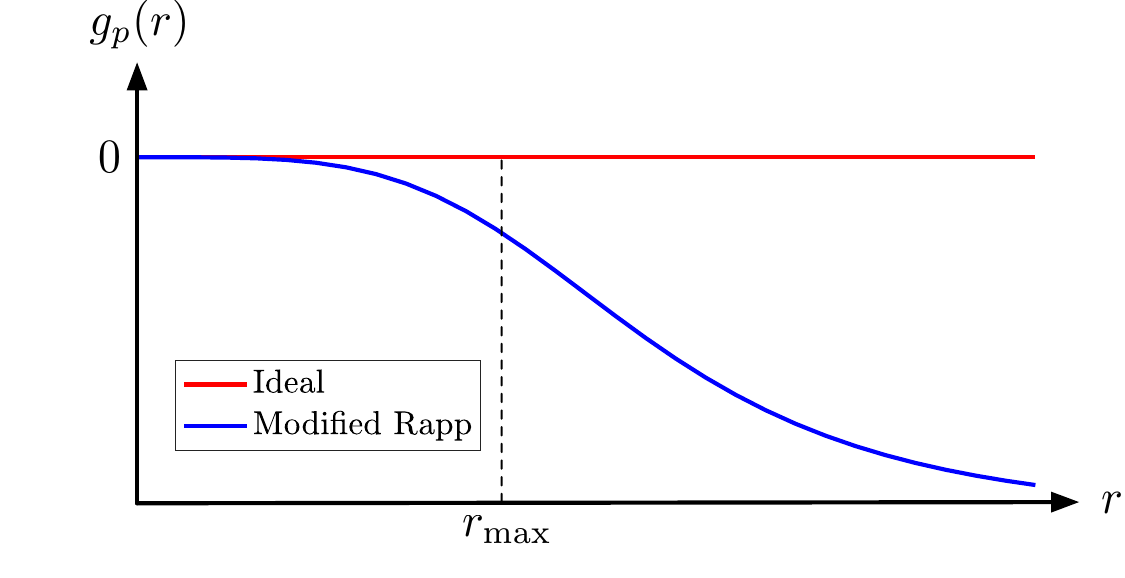}}
\caption{Illustration of the ideal and modified Rapp PA models.}
\label{fig:PA_response}
\end{figure}

Let us start with introducing the PA model.
Consider the widely used memoryless PA model~\cite{schreurs2008rf,ochiai2013analysis}
\begin{equation}\label{eq:PA_general}
	x_{\rm out}(t) = G(x_{\rm in}(t)),
\end{equation}
where
$x_{\rm out}(t)$ and $x_{\rm in}(t)$  are the  complex baseband  PA output and input  signals, respectively;
$G(\cdot)$ is the PA response function, which is assumed to take the form
\[
G(x) = g_a(|x|) e^{\jj \cdot {\rm arg}(x)} e^{\jj \cdot g_p(|x|)},
\]
where $g_a(\cdot)$ and $g_p(\cdot)$ are the amplitude and phase responses, also referred to as
the amplitude-to-amplitude (AM-AM) and amplitude-to-phase (AM-PM) conversions, respectively.

In the ideal case, the PA output should be
a linear amplification of the PA input within the allowable PA output amplitude range.
To be more specific, we desire a linear PA response up to the maximum PA output amplitude, i.e.,
\begin{equation}\label{eq:ideal_pa}
	g_a(r) =
	\begin{cases}
		Ar, & 0 \le r \le r_{\rm max},\\
		Ar_{\rm max}, & r > r_{\rm max},
	\end{cases}
	\quad
	g_p(r) = 0,
\end{equation}
where $A$ is the PA gain, and $Ar_{\rm max}$ is the maximum PA output amplitude, with $r_{\rm max}$ being a reference PA input amplitude;
see Fig.~\ref{fig:PA_response} for an illustration.
However, due to the physics of semiconductors, realistic PAs generally exhibit nonlinear responses.
In the literature, there are several mathematical models to describe $g_a(\cdot)$ and $g_p(\cdot)$ for realistic PAs~\cite{schreurs2008rf,ochiai2013analysis,3gpp_pa,saleh1981frequency,rapp1991effects}.
A popular model is the modified Rapp model, which was proposed in 3GPP
for performance evaluation in 5G systems~\cite{3gpp_pa}.
Specifically, the modified Rapp model is expressed as
\begin{equation}\label{eq:rapp}
	g_a(r) \!= \!\frac{Ar}{\left( 1+\left( r/r_{\rm max} \right)^{2\varphi} \right)^{\frac{1}{2\varphi}}}, ~
	g_p(r) \!= \! \frac{B  r^\zeta}{1+\left({r}/{C}\right)^\zeta}~ ({\rm rad}),
\end{equation}
where $\varphi$, $\zeta$, $B$ and $C$ are the fitting parameters;
see Fig.~\ref{fig:PA_response}.
We see that under such a nonlinear PA model, the PA output is distorted by the PA nonlinearity.
Apart from the modified Rapp model,  other commonly used models include the traveling-wave tube amplifier (TWTA) model~\cite{saleh1981frequency} and the solid state power amplifier (SSPA) model~\cite{rapp1991effects}.

\subsection{Prior Art}
\label{sec:prior_art}

In this subsection, we review two conventional approaches to combating the PA distortions.
The first one is {\it power back-off}~\cite{schenk2008rf,jedda2017precoding,spano2017symbol}.
This method is based on a basic observation of realistic PA responses---the PA response is near-linear for low PA input amplitude;
the modified Rapp model shown in Fig.~\ref{fig:PA_response} is an example.
The idea is to reduce the PA input amplitude such that the PA response is near-linear.
A common way of performing power back-off is to
shrink the PA input amplitude $r$ such that it satisfies
\begin{equation}\label{eq:1db}
	r \le r_{\rm 1dB},
\end{equation}
where $r_{\rm 1dB}$ is the {\it 1dB compression point}, usually recognized as the border between the linear and nonlinear regions of the PA response~\cite{ochiai2013analysis}.
More precisely, $r_{\rm 1dB}$ is defined as the PA input amplitude, where the difference of the output amplitudes between the linear and the PA response is 1dB, i.e.,
\[
20\log_{10}\frac{Ar_{\rm 1dB}}{g_a(r_{\rm 1dB})} = 1;
\]
see the example in Fig.~\ref{fig:PA_response_am}.
The power back-off method is effective in PA distortion reduction and also simple in implementation,
but such a direct power back-off strategy results in reduced power efficiency~\cite{ochiai2013analysis}.

\begin{figure}[t]
	\centering
	\includegraphics[width=0.6\linewidth]{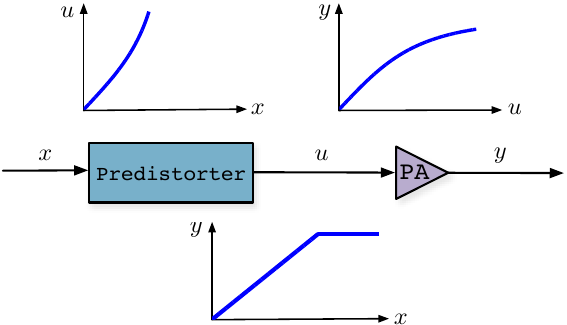}
	\caption{The principle of DPD.}
	\label{fig:PD}
\end{figure}

\begin{figure}
	\centering
	\includegraphics[width=0.6\linewidth]{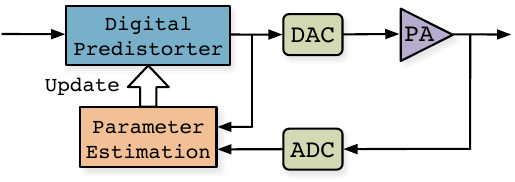}
	\caption{Block diagram of a general DPD system.}
	\label{fig:DPD}
\end{figure}

The second approach is {\it digital predistortion} (DPD)~\cite{ghannouchi2009behavioral,guan2014green,katz2016evolution},
which is currently the most widely applied PA linearization method in wireless BSs.
As illustrated in Fig.~\ref{fig:PD},
DPD works by adding
a nonlinear component, called a predistorter, in front of the nonlinear PA, such that the combined response
of the predistorter and the PA
is an ideal PA response in \eqref{eq:ideal_pa}.
Fig.~\ref{fig:DPD} shows how DPD is realized in practice.
For ease of implementation, the predistorter is implemented in digital domain, and the digital predistorter needs to be continuously adjusted to accommodate the PA response change with time.

Ideally, DPD can achieve excellent PA linearization performance,
but the challenges lie in hardware implementations.
To be specific, DPD requires
i) an expensive high-resolution and fast analog-to-digital converter (ADC) at the feedback loop to capture the PA distortion behavior;
ii) an accurate parametric model for the PA transfer function;
iii) a fast training algorithm to update the PA model parameters in real time.
And we have such requirements for each of the PAs.
Therefore, the hardware complexity of DPD can be high in practice, especially in massive MIMO systems with numerous PAs.

\section{Spatial $\Sigma \Delta$ Modulation for PA Distortion Mitigation}
\label{sec:sd_all}

In this section, we present our spatial $\Sigma \Delta$ modulation concept for handling the PA distortion problem.
The idea comes from one-bit $\Sigma \Delta$ noise shaping, which was developed to reduce the quantization noise effects in one-bit MIMO precoding~\cite{shao2019one}.
To explain the idea, we first review the principle of one-bit $\Sigma \Delta$ noise shaping and then describe how we adopt the principle to combat the PA distortion effects.

\subsection{Review of One-Bit ${\Sigma \Delta}$ Noise Shaping}
\label{sec:review_SD}

\begin{figure}
	\centering
	\includegraphics[width=0.5\linewidth]{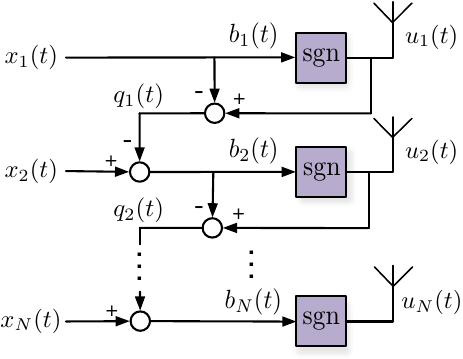}
	\caption{The first-order ${\Sigma \Delta}$ modulator for quantization noise shaping.}
	\label{fig:sd_onebit}
\end{figure}

We review the spatial $\Sigma\Delta$ modulation in \cite{shao2019one}, which was proposed to address the quantization effect caused by one-bit digital-to-analog converters (DACs).
Consider the massive MIMO downlink scenario, where a BS with $N$ antennas transmits signals to multiple users and each transmit antenna is equipped with one-bit DACs for low-cost implementation.
To let the reader see the idea easier, we describe the principle of spatial $\Sigma\Delta$ modulation through real-valued continuous-time transmitted signals, rather than through complex-valued discrete-time transmitted signals as in the original paper~\cite{shao2019one}.
Fig.~\ref{fig:sd_onebit} shows the first-order $\Sigma \Delta$ modulator structure,
where
$x_n(t) \in \Rbb$ is the modulator input,
$b_n(t) \in \Rbb$ is the quantizer input,
${\rm sgn}(\cdot)$ is the signum (or one-bit quantization) function,
$u_n(t) = {\rm sgn}(b_n(t))$ is the quantizer output,
and $q_n(t) \in \Rbb$ is the quantization noise.
We see in Fig.~\ref{fig:sd_onebit} that the input-output relation is
\begin{equation}\label{eq:sd_unt}
u_n(t) = x_n(t) +  q_n(t) -  q_{n-1}(t),
\end{equation}
for $n=1,\dots,N$, where $q_0(t) = 0$.
As a key assumption in one-bit $\Sigma \Delta$ noise shaping, the transmit antennas are assumed to be arranged as a uniform linear array (ULA).
Let
\begin{equation}\label{eq:angular_response}
	\ba(\theta) = \left[ 1,e^{-\jj\frac{2\pi d}{\lambda}\sin(\theta)}, \dots, e^{-\jj(N-1)\frac{2\pi d}{\lambda}\sin(\theta)} \right]^\Tsf
\end{equation}
be the angular response of the ULA,
where
$\theta \in [-\pi/2,\pi/2]$ is the angle,
$\lambda$ is the carrier wavelength
and $d \le {\lambda}/{2}$ is the inter-antenna spacing.
Assuming unit channel gain, we can express the signal received at angle $\theta$ as
\begin{align}
	\ba(\theta)^\Tsf \bu(t) &= \ba(\theta)^\Tsf \bx(t) + \xi_\omega(t), \notag\\
	\xi_\omega(t) &= (1 - e^{-\jj \omega})  Q_\omega(t)  +  q_N(t) e^{- \jj \omega(N-1)}, \label{eq:xi_onebit}
\end{align}
where
$\bu(t) \!\!=\!\! [u_1(t),\dots,u_N(t)]^\Tsf$;
$\bx(t) \!\!=\!\! [x_1(t),\dots,x_N(t)]^\Tsf$;
$\omega \!=\! \frac{2\pi d}{\lambda} \sin(\theta)$ denotes the spatial frequency associated with $\theta$;
$\xi_\omega(t)$ is the received quantization noise with
$Q_\omega(t) \!=\! \sum_{n=1}^{N-1} \! q_n(t) e^{-\jj \omega(n-1)}$.
We observe from~\eqref{eq:xi_onebit} that the quantization noise term $Q_\omega(t)$ is shaped by a high-pass response $(1 - e^{-\jj \omega})$,
and hence the shaped noise $(1 - e^{-\jj \omega})  Q_\omega(t)$ is expected to be high-frequency noise.
The noise power should decrease as $|\omega|$ decreases, and $|\omega|$ decreases with $|\theta|$ and $d$.
Therefore, to mitigate the quantization noise effects, it is desired to
\begin{enumerate}[i)]
\item
keep the operating range of user angles $\theta$ small, e.g., $\theta \in [-30^\circ, 30^\circ]$;
\item
use a small inter-antenna spacing $d$, e.g., $d = \lambda/8$.
\end{enumerate}
It is worth noting that the item i) corresponds to a sectored antenna array scenario;
and that, for ii), we cannot make $d$ arbitrarily small due to physical antenna size and mutual coupling effects.

There may be stability problems with the $\Sigma \Delta$ modulator in Fig.~\ref{fig:sd_onebit}.
Due to the feedback loop, the quantization noise amplitude $|q_n(t)|$ may be accumulated with $n$ and grow large.
In some extreme cases, we may have $|q_{n}(t)| \to \infty$ as $n \to \infty$.
This phenomenon is referred to as {\it overloading} in the $\Sigma \Delta$ literature.
Overloading can be avoided by appropriately constraining the modulator input amplitudes.

\begin{Fact} \label{fact:no_overloading_onebit}
	{\bf (see, e.g.,~\cite{gray1990quantization,shao2019one})}
	Consider the first-order $\Sigma \Delta$ modulator in Fig.~\ref{fig:sd_onebit}.
	If $|x_n(t)| \le 1$ for all $n$, then it holds that $|q_n(t)| \le 1$ for all $n$.
\end{Fact}

\noindent
Under the no-overloading condition, i.e., $|x_n(t)| \le 1$ for all $n$,
it is common to make the following assumption.
\begin{Asm}\label{asm:q_onebit}
	Consider the first-order $\Sigma \Delta$ modulator in Fig.~\ref{fig:sd_onebit}.
	Under the condition that $|x_n(t)| \le 1$ for all $n$, the quantization noise $q_{n}(t)$ is uniformly distributed on $[-1,1]$ and is independently and identically distributed (i.i.d.) over $n$.
	Also, each $q_{n}(t)$ is independent of any other random variables.
\end{Asm}

\noindent
Under Assumption~\ref{asm:q_onebit}, the power of the received quantization noise $\xi_\omega(t)$ in~\eqref{eq:xi_onebit} is calculated by
\[
\mathbb{E}[|\xi_\omega(t)|^2] =
\frac{4(N-1)}{3}   \sin^2 \left( \frac{\pi d}{\lambda} \sin(\theta)\right) + \frac{1}{3},
\]
which is seen to reduce as $|\theta|$ and/or $d$ decreases.

\subsection{${\Sigma \Delta}$ PA Distortion Shaping}
\label{sec:sd}

\begin{figure}[t]
	\centering
	\includegraphics[width=0.5\linewidth]{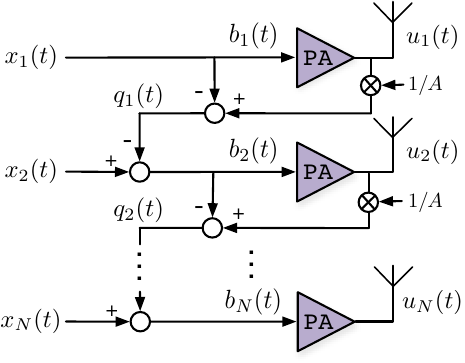}
	\caption{The first-order ${\Sigma \Delta}$ modulator for PA distortion shaping.}
	\label{fig:sd}
\end{figure}

We now adopt the principle of one-bit $\Sigma \Delta$ noise shaping
to handle the PA nonlinear distortions.
The idea is to spatially shape the PA distortions in the same way as how we shape the quantization noise in the previous subsection.
The first-order $\Sigma \Delta$ modulator structure for PA distortion shaping is shown in Fig.~\ref{fig:sd},
where
$x_n(t) \in \Cbb$ is the modulator input;
$b_n(t) \in \Cbb$ is the PA input;
$u_n(t) = G(b_n(t))$ is the PA output, with $G(\cdot)$ being the PA response function;
$A$ is the PA gain;
$q_n(t) \in \Cbb$ is the PA distortion.
Assuming a ULA at the BS
and following similar derivations as in \eqref{eq:sd_unt}--\eqref{eq:xi_onebit},
the signal received at angle $\theta$ is given by
\begin{align}
	\ba(\theta)^\Tsf \bu(t) &= A\ba(\theta)^\Tsf \bx(t) + \xi_\omega(t), \label{eq:au} \\ 
	\xi_\omega(t) &= A(1 - e^{-\jj \omega})  Q_\omega(t)  +  Aq_N(t) e^{- \jj \omega(N-1)}, \label{eq:xi}
\end{align}
where $\xi_\omega(t)$ now represents the $\Sigma \Delta$-shaped PA distortion;
$\omega$ and $Q_\omega(t)$ are defined in the same way as in~\eqref{eq:xi_onebit}.
The PA distortion term $A(1 - e^{-\jj \omega})  Q_\omega(t)$ can be seen as high spatial frequency noise.
By operating in a low-angle region for the users and by using a small inter-antenna spacing $d$, we can mitigate the effects of the PA distortions on the received signals---same as what happens in one-bit $\Sigma \Delta$ noise shaping.
In PA distortion shaping, the no-overloading condition is characterized as follows.
\begin{Fact} \label{Fac:no_overloading}
	Consider the first-order $\Sigma \Delta$ modulator in Fig.~\ref{fig:sd}.
	Given any $\chi > 0$, define
	\begin{equation}\label{eq:C}
	\psi \triangleq \max_{|z| \le \chi}|G(z)/A - z|.
	\end{equation}
	If the modulator input amplitudes are constrained by
	\begin{equation}\label{eq:cont_no_overloading}
		|{x}_{n}(t)| \le   \chi - \psi,
	\end{equation}
    for all $n$, then it holds that $|b_n(t)| \le \chi$ and $|q_{n}(t)| \le \psi$ for all $n$.
\end{Fact}

\noindent
The proof of Fact~\ref{Fac:no_overloading} is relegated to the Appendix.
The variable $\psi$ in~\eqref{eq:C} describes the largest PA distortion under the PA input amplitude range $|z| \le \chi$.
It remains to specify the choice of $\chi$.
Intuitively speaking, when $\chi$ increases, the signal power tends to increase as $|b_n(t)| \le \chi$;
on the other hand, the PA distortion power also tends to increase as $|q_{n}(t)| \le \psi$ and $\psi$ scales with $\chi$.
Therefore, there is a tradeoff.
In this work, we heuristically choose $\chi = r_{\rm max}$ and it will be shown in our numerical study that such choice empirically works well.
In the same spirit as Assumption~\ref{asm:q_onebit} in the one-bit DAC case,
we make the following assumption.
\begin{Asm}\label{asm:q}
	Consider the first-order $\Sigma \Delta$ modulator in Fig.~\ref{fig:sd}.
	Under the condition in~\eqref{eq:cont_no_overloading}, the PA distortion $q_{n}(t)$ is i.i.d. over $n$ with $|q_{n}(t)| \sim \mathcal{U}{[0,\psi]}$ and $\arg(q_{n}(t)) \sim \mathcal{U}{[-\pi,\pi]}$.
	Also, each $q_{n}(t)$ is independent of any other random variables.
\end{Asm}

\noindent
Under Assumption~\ref{asm:q}, the power of the $\Sigma \Delta$-shaped PA distortion $\xi_\omega(t)$ in~\eqref{eq:xi} can be shown to be
\[
\mathbb{E}[|\xi_\omega(t)|^2] =
\frac{4(N-1)A^2\psi^2}{3}   \sin^2 \left( \frac{\pi d}{\lambda} \sin(\theta)\right) + \frac{A^2\psi^2}{3}.
\]
We see that the shaped distortion power reduces as $|\theta|$ and/or $d$ decreases.

\subsection{A Tail-Removing ${\Sigma \Delta}$ Scheme}
\label{sec:sd_tr}

It is observed from \eqref{eq:xi} that the PA distortion $q_N(t)$ at the last antenna cannot be shaped by the first-order $\Sigma \Delta$ modulator.
By our empirical experience, $q_N(t)$ could have a non-negligible effect on the received signals.
To overcome this drawback, we suggest to employ a linear PA at the last antenna to make $q_N(t)=0$ in \eqref{eq:xi}.
The hardware complexity of employing a high-quality linear PA  at only one antenna should be affordable in practice.
In this case, the received signal at angle $\theta$ becomes
\begin{equation}\label{eq:receive_sig_sd}
\begin{aligned}
\ba(\theta)^\Tsf \bu(t) &= A\ba(\theta)^\Tsf \bx(t) + \xi_\omega(t), \\
\xi_\omega(t) &= A(1 - e^{-\jj \omega})  Q_\omega(t),
\end{aligned}
\end{equation}
where all the PA distortions are shaped.
In addition, under Assumption~\ref{asm:q}, the shaped distortion power is
\[
\mathbb{E}[|\xi_\omega(t)|^2] \!= \!
\frac{4(N-1)A^2\psi^2}{3}   \sin^2 \left( \frac{\pi d}{\lambda} \sin(\theta)\right).
\]
We will refer to the above scheme as the Tau-Sigma-Delta (T$ \Sigma \Delta$) scheme in the sequel.

\begin{figure}[t]
	\centering
	\includegraphics[width=0.5\linewidth]{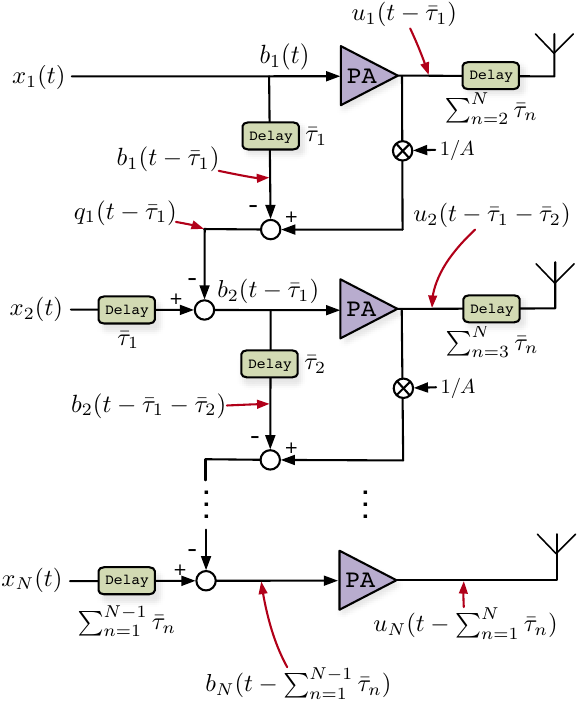}
	\caption{ The first-order ${\Sigma \Delta}$ modulator with time-delay components.}
	\label{fig:sd_delay}
\end{figure}

\subsection{Further Discussions}

In this subsection we discuss some practical aspects with the $\Sigma \Delta$ PA distortion shaping concept presented in the previous subsections.
In practice, the $\Sigma \Delta$ modulator is implemented in analog domain.
The issue arising is that the PAs may introduce certain time delays to the signals.
Let $\bar \tau_n$ be the PA time delay at the $n$th antenna.
Then the delayed PA output feedback
$u_n(t-\bar \tau_n)/A$
and the PA input $b_n(t)$ will be mismatched.
This issue can be fixed by adding time-delay components in appropriate positions;
see Fig.~\ref{fig:sd_delay} for the structure.
Note that such structure is realizable in practice.
For example, in the traditional feedforward PA linearization method~\cite{cripps2006rf}, a similar structure with time-delay components has been realized.

\begin{figure}[t]
	\centering
	\includegraphics[width=0.5\linewidth]{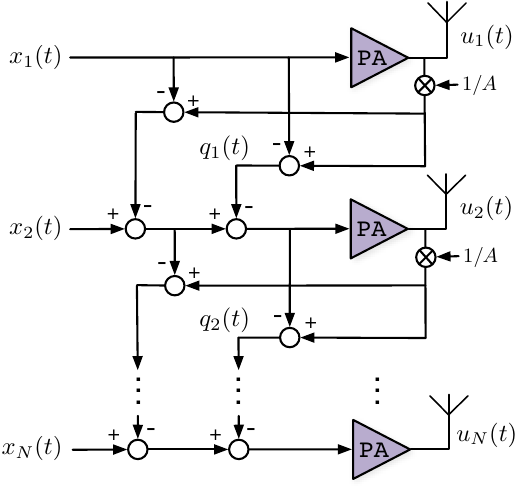}
	\caption{The second-order ${\Sigma \Delta}$ modulator.}
	\label{fig:sd_second_order}
\end{figure}

It is also worthwhile to discuss other $\Sigma \Delta$ modulator structures.
Conceptually, the first-order $\Sigma \Delta$ modulator in Fig.~\ref{fig:sd} can be extended to higher-order $\Sigma \Delta$ modulators,
which aim to provide stronger PA distortion shaping effects~\cite{aziz1996overview,corey2016spatial,shao2020multiuser}.
For example, we can consider the second-order $\Sigma \Delta$ modulator  illustrated in Fig.~\ref{fig:sd_second_order}.
One can show that for this second-order extension, the T$\Sigma \Delta$ scheme applies linear PAs at the last two antennas and has the received signal given by
\begin{equation}\label{eq:receve_sig_second_sd}
\ba(\theta)^\Tsf \bu(t) = A\ba(\theta)^\Tsf \bx(t) + A(1 - e^{-\jj \omega})^2  \tilde Q_\omega(t),
\end{equation}
where $\tilde Q_\omega(t)=\sum_{n=1}^{N-2}q_n(t)e^{- \jj \omega(n-1)}$.
Compared with the first-order scheme in~\eqref{eq:receive_sig_sd}, the PA distortion shaping filter in~\eqref{eq:receve_sig_second_sd} is of higher order and has a sharper shape,
and the PA distortions may be more well suppressed in the low-angle region.
Also, the no-overloading condition is given by $|x_n(t)| \le \chi - 3\psi$ for all $n$.
But we should also note that the second-order $\Sigma \Delta$ modulator  requires higher hardware complexity to realize.

\section{Application to Multi-User MIMO-OFDM Precoding}
\label{sec:application}

In this section, we apply our spatial $\Sigma \Delta$ modulation concept to the multi-user MIMO-OFDM downlink scenario.
The system model under consideration is depicted in Fig.~\ref{fig:sys_mod}, where
a BS equipped with $N$ antennas simultaneously serves $K$ single-antenna users over a frequency-selective channel.
The modulation scheme is OFDM.
The transmit antennas are assumed to have nonlinear PA effects, and the first-order $\Sigma \Delta$ modulator in Fig.~\ref{fig:sd} is incorporated to combat the PA nonlinear distortion effects.

\subsection{OFDM Signal Model}

\begin{figure*}[t]
	\centering
	\includegraphics[width=0.9\linewidth]{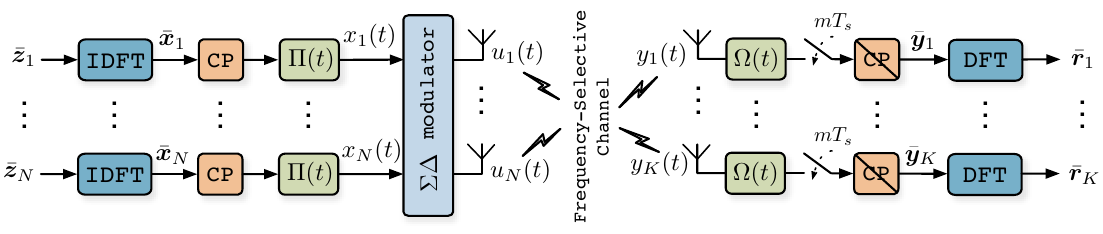}
	\caption{Multi-user MIMO-OFDM system model with the first-order ${\Sigma \Delta}$ modulator.}
	\label{fig:sys_mod}
\end{figure*}

To begin with, we describe the implementation of the OFDM signals $x_1(t), \dots, x_N(t)$ in Fig.~\ref{fig:sys_mod}.
For simplicity, we focus on one OFDM block.
The ideal OFDM signal we desire to implement takes the form
\begin{equation}\label{eq:OFDM_sig_ideal}
	x^{\sf ideal}_n(t) =  \sum_{p=0}^{M_s-1} z_{n,p} e^{\jj  \frac{2\pi p}{T}t}, \quad  0 \le t < T,
\end{equation}
for $n=1,\dots,N$,
where
$M_s$ is the number of subcarriers,
$z_{n,p} \in \Cbb$ is the signal transmitted at antenna $n$ and subcarrier $p$,
and $T$ is the duration of one OFDM symbol.
Before transmission, a cyclic prefix (CP) is inserted at the beginning of each $x^{\sf ideal}_n(t)$ to avoid inter-symbol interference caused by the frequency-selective channel, i.e.,
\begin{equation}\label{eq:OFDM_sig_ideal_cp}
	x^{\sf ideal}_n(t) = x^{\sf ideal}_n(t+T), \quad -T_{\sf CP} \le t < 0,
\end{equation}
where $T_{\sf CP}$ is the duration of the CP.

\begin{figure}[t]
	\centering
	\includegraphics[width=0.6\linewidth]{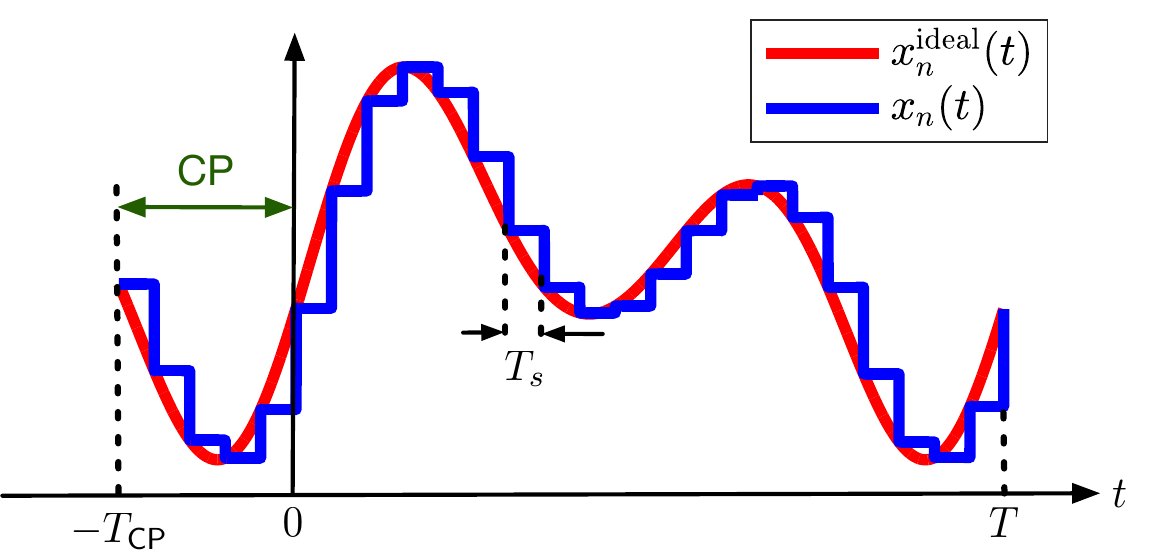}
	\caption{Illustration of the OFDM signal implementation.}
	\label{fig:OFDM_approx}
\end{figure}

In accordance with the LTE and 5G standards~\cite{dahlman20134g,5G_ofdm}, we implement $x^{\sf ideal}_n(t)$ via inverse discrete Fourier transform (IDFT) with oversampling.
As illustrated in Fig.~\ref{fig:OFDM_approx}, consider approximating $x^{\sf ideal}_n(t)$ in \eqref{eq:OFDM_sig_ideal}--\eqref{eq:OFDM_sig_ideal_cp} by
\begin{equation}\label{eq:OFDM_sig}
	x_n(t) =  \sum_{m=-M_{\sf CP}}^{M-1}  x_{n,m} \Pi(t-mT_s), \quad -T_{\sf CP} \le t < T,
\end{equation}
where
$x_{n,m} = x_n^{\sf ideal}(mT_s)$ is a sample of $x_n^{\sf ideal}(t)$;
$T_s = T/(\kappa M_s)$ is the sampling period, with $\kappa \ge 1$ being the oversampling ratio;
$M_{\sf CP} = T_{\sf CP}/T_s$ is the number of samples for the CP;
$M = \kappa M_s$;
$\Pi(t)$ is a rectangular pulse with width $T_s$, i.e.,
\begin{equation*}
	\Pi(t) =
	\begin{cases}
		1, &\quad 0 \le t < T_s,\\
		0, &\quad {\rm otherwise}.
	\end{cases}
\end{equation*}

\noindent
We implement the OFDM signals by generating the approximate signals in~\eqref{eq:OFDM_sig};
the approximation accuracy improves as the oversampling ratio $\kappa$ increases.
In practice, equation~\eqref{eq:OFDM_sig} can be realized by DACs with sample-and-hold operations.
Moreover, according to the equations in~\eqref{eq:OFDM_sig_ideal}--\eqref{eq:OFDM_sig_ideal_cp}, the samples $x_{n,m}$'s in~\eqref{eq:OFDM_sig} are given by
\begin{subnumcases}{x_{n,m} =}
	\sum_{p=0}^{M_s-1} z_{n,p} e^{\jj  \frac{2\pi mp}{M}},   & $0 \le m \le M-1,$ \label{eq:idft_scalar} \\
	x_{n,m+M},   &$-M_{\sf CP} \le m \le -1.$ \label{eq:cp_add}
\end{subnumcases}
Equation~\eqref{eq:idft_scalar} can be rewritten in an equivalent vector form
\begin{equation}\label{eq:idft_vect}
	[x_{n,0},\dots,x_{n,M-1}]^\Tsf = \bF [z_{n,0},\dots,z_{n,M_s-1},0,\dots,0]^\Tsf,
\end{equation}
where
$\bF = [e^{\jj \frac{2\pi mp}{M}}]_{m,p} \in \Cbb^{M \times M}$ is a scaled IDFT matrix.
Equation~\eqref{eq:idft_vect} indicates that the $x_{n,m}$'s can be obtained by performing IDFTs to appropriately zero-padded $z_{n,p}$'s.
For convenience, we will represent~\eqref{eq:idft_vect} as
\begin{equation}\label{eq:bar_bx_n}
	\bar \bx_n = \bF_s \bar \bz_n,
\end{equation}
where
$\bar \bx_n \!\!=\!\! [x_{n,0},\dots,x_{n,M-1}]^\Tsf$, $\bar \bz_n \!\!=\!\! [z_{n,0},\dots,z_{n,M_s-1}]^\Tsf$, and
$\bF_s \in \Cbb^{M \times M_s}$ aggregates the first $M_s$ columns of $\bF$.
The left-hand side of Fig.~\ref{fig:sys_mod} illustrates the aforementioned implementation process that involves the IDFT operation in \eqref{eq:bar_bx_n}, the CP addition operation in \eqref{eq:cp_add} and the signal generation operation in \eqref{eq:OFDM_sig}.


\subsection{Transmission Signal Model}

\begin{figure}[t]
	\centering
	\includegraphics[width=0.8\linewidth]{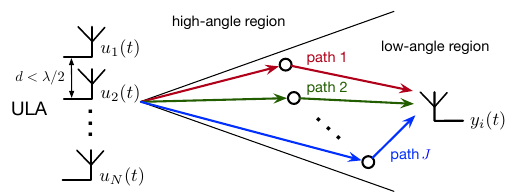}
	\caption{The multi-path channel.}
	\label{fig:multi_path}
\end{figure}

The OFDM signals $x_n(t)$'s are then fed into the first-order $\Sigma \Delta$ modulator.
We see from Fig.~\ref{fig:sd} that the $\Sigma \Delta$ modulator output is given by
\begin{equation}\label{eq:sys_mod_unt}
	u_n(t) = Ax_n(t) +  A(q_n(t) -  q_{n-1}(t)).
\end{equation}
Following the $\Sigma \Delta$ PA distortion shaping rationale in Section~\ref{sec:sd_all}, we assume a ULA at the BS operating in a low-angle region and using a small inter-antenna spacing.
We also assume that the channel from the BS to each user has $J$ paths;
see Fig.~\ref{fig:multi_path} for the scenario.
The channel input-output relation is described by

\begin{equation}\label{eq:cont_channel_mod}
	{y}_i(t) =  \sum_{j=1}^J \alpha_{i,j} \ba(\theta_{i,j})^\Tsf {\bu}(t - \tau_{i,j}) +  v_i(t), 
\end{equation}
where
${y}_i(t)$ is the received signal of user $i$;
$\alpha_{i,j}$,  $\theta_{i,j}$ and $\tau_{i,j}$ are the gain, angle and delay of path $j$ to user $i$, respectively;
$\ba(\theta)$ is the angular response of the ULA defined in~\eqref{eq:angular_response};
$v_i(t)$ is background noise.

At the user side,
the received signal $y_i(t)$ is low-pass filtered and then sampled to produce the discrete-time received signal
\begin{equation}\label{eq:yim}
	y_{i,m} = (y_i \otimes \Omega)(mT_s), 
\end{equation}
where
$\otimes$ denotes the convolution operator and
$\Omega(t)$ is the impulse response of the receive low-pass filter.
Combining \eqref{eq:OFDM_sig}, \eqref{eq:sys_mod_unt}, \eqref{eq:cont_channel_mod} and \eqref{eq:yim} gives the equivalent discrete-time system model
\begin{equation}\label{eq:dist_time_mod}
y_{i,m} = \sum_{\ell=0}^{L-1} \bh_{i,\ell}^\Tsf  \bx_{m-\ell} + \xi_{i,m} +  v_{i,m},
\end{equation}
where
$\bx_m = [x_{1,m},\dots,x_{N,m}]^\Tsf$;
$L$ is the number of channel taps;
$
\bh_{i,\ell} = A\sum_{j=1}^J \alpha_{i,j} \ba(\theta_{i,j})  (\Pi \otimes \Omega)(\ell T_s - \tau_{i,j});
$
$\xi_{i,m}$ is the received PA distortion noise, expressed as
\begin{equation}\label{eq:xiim}
\xi_{i,m} =   \sum_{j=1}^J   (1-e^{-\jj \omega_{i,j}})  \sum_{n=1}^{N-1} \alpha_{i,j} \hat q_{n,m}^{i,j}   +  \sum_{j=1}^J \alpha_{i,j}\hat q_{N,m}^{i,j}
\end{equation}
with $\hat q_{n,m}^{i,j} = A  e^{-\jj \omega_{i,j}(n-1)} (q_n \otimes \Omega)(mT_s-\tau_{i,j})$ and
$\omega_{i,j} = \frac{2\pi d}{\lambda} \sin(\theta_{i,j})$;
$v_{i,m}$ is background noise and we assume $v_{i,m} \sim \mathcal{CN}(0,\sigma_v^2)$~\cite{tse2005fundamentals}.
We see from~\eqref{eq:xiim} that each distortion term $\sum_{n=1}^{N-1} \alpha_{i,j} \hat q_{n,m}^{i,j}$ is shaped by a high-pass filter in space.

%
%
%
%

Finally, the frequency-domain received signals are obtained by performing CP removal and discrete Fourier transform (DFT) to the $y_{i,m}$'s, i.e.,
\begin{equation}\label{eq:bri}
	\bar \br_i = \bF_s^\Hsf \bar \by_i,
\end{equation}
where
$\bar \br_i  \!\! = \!\! [r_{i,0}, \dots, r_{i,M_s-1}]^\Tsf$ and
$\bar \by_i \!\! = \!\! [y_{i,0}, \dots, y_{i,M-1}]^\Tsf$ are the frequency-domain and CP-free discrete-time received signals for user $i$, respectively.
By combining~\eqref{eq:bar_bx_n}, \eqref{eq:dist_time_mod} and \eqref{eq:bri}, we arrive at the frequency-domain system model
\begin{equation}\label{eq:freq_mod}
	r_{i,p} = \check \bh_{i,p}^\Tsf \bz_p+ \check \xi_{i,p} + \check v_{i,p}, 
\end{equation}
where
$\check \bh_{i,p} = \sum_{\ell=0}^{L-1} {\bh}_{i,\ell} e^{-\jj \frac{2 \pi \ell (p-1)}{M} }$ is the channel for user $i$ at subcarrier $p$,
$\bz_p = [z_{1,p},\dots,z_{N,p}]^\Tsf$ is the transmitted signal at subcarrier $p$,
$[\check \xi_{i,0}, \dots, \check \xi_{i,M_s-1}]^\Tsf = \bF_s^\Hsf [\xi_{i,0},\dots,\xi_{i,M-1}]^\Tsf$ is the received PA distortion noise, and
$\check v_{i,p} \sim \mathcal{CN}(0,\sigma^2_v)$ is background noise.


\subsection{MIMO-OFDM Precoding under Signal Amplitude Constraints}
\label{sec:precoding}

To complete the picture, we design the multi-user precoding under the effective model \eqref{eq:freq_mod}.
To describe,
let $s_{i,p} \in \Cbb$ be the data symbol for user $i$ at subcarrier $p$.
We assume that each $s_{i,p}$ is drawn from a quadrature amplitude modulation (QAM) constellation, i.e.,
\begin{equation*}\label{eq:QAM_cons}
	s_{i,p} \in \setS = \{ s_R + \jj s_I  \mid  s_R, s_I \in \{ \pm 1, \pm 3, \ldots, \pm (2D - 1) \} \},
\end{equation*}
where $D$ is a positive integer (the QAM size is $4D^2$).
Assuming perfect knowledge of the channels $\check \bh_{i,p}$'s at the BS, we aim to design the transmitted signals $\bz_p$'s such that the users will receive their intended data symbols.
More specifically, we want the noise free part of $r_{i,p}$ in~\eqref{eq:freq_mod} to take the form
\begin{equation}\label{eq:symbol_shaping}
	\check \bh_{i,p}^\Tsf \bz_p \approx \beta_i s_{i,p},
\end{equation}
where $\beta_i>0$ denotes a QAM constellation scaling factor for user $i$.
With a larger $\beta_i$, the received signal can be more robust to noise.
The $\beta_i$'s are designed by the BS and each user is assumed to know its $\beta_i$.
As a new design requirement introduced by the spatial $\Sigma \Delta$ approach, the no-overloading condition in~\eqref{eq:cont_no_overloading} must be guaranteed.
It follows from the temporal signal model~\eqref{eq:OFDM_sig} that the no-overloading condition~\eqref{eq:cont_no_overloading} amounts to constraining the discrete-time signals $\bX = [\bx_0, \dots, \bx_{M-1}]$ by
\begin{equation}\label{eq:no_overloading}
\bX \in \setX = \{ \bX \in \Cbb^{N \times M} ~\vert~	\|\bX\|_{\rm max} \le \chi - \psi \},
\end{equation}
where
$\|\bX\|_{\rm max} = \max_{i,j}|x_{i,j}|$.

\subsubsection{$\Sigma \Delta$ Zero-Forcing}

We first develop a zero-forcing (ZF) scheme for the precoding design.
In the ZF scheme, the transmitted signals are designed by the form
\begin{equation}\label{eq:zf}
	\bz_p =
	\frac{1}{\Gamma} \check \bH_p^\dagger \bs_p, \quad p=0,\dots,M_s-1,
\end{equation}
where
$\check \bH_p^\dagger = \check \bH_p^\Hsf (\check \bH_p \check \bH_p^\Hsf)^{-1}$,
$\check \bH_p = [\check \bh_{1,p},\dots,\check \bh_{K,p}]^\Tsf$, $\bs_p = [s_{1,p},\dots,s_{K,p}]^\Tsf$, and $\Gamma$ is a normalization factor to ensure that the no-overloading condition in~\eqref{eq:no_overloading} holds.
It is easy to see that such $\Gamma$ is given by
\[
\Gamma = \frac{\| [\check\bH_0^\dagger\bs_0,\dots,\check\bH_{M_s-1}^\dagger\bs_{M_s-1}] \bF_s^\Tsf \|_{\rm max} }{\chi - \psi}.
\]
The noise-free received signal is then expressed as
\begin{equation*}
	\check \bh_{i,p}^\Tsf  \bz_p = \frac{1}{\Gamma} s_{i,p},
\end{equation*}
from which we observe that the corresponding QAM constellation scaling factors are
$
\beta_i = {1}/{\Gamma},~ \forall i.
$


\subsubsection{$\Sigma \Delta$ Symbol-Level Precoding}

We next develop a symbol-level precoding (SLP) scheme.
The idea is to jointly design the transmitted signals $\bz_p$'s and the scaling factors $\beta_i$'s to optimize the detection probability (DP) of all data symbols.
We begin with characterizing the effective PA distortion noise power
$\mathbb{E}[|\xi_{i,m}|^2]$ for $\xi_{i,m}$ in \eqref{eq:dist_time_mod}.
By noting the difficulty of exactly calculating $\mathbb{E}[|\xi_{i,m}|^2]$, we resort to convenient approximation.
By invoking Fact~\ref{Fac:no_overloading}, we see that under the no-overloading condition,
it holds that $|q_n(t)| \le \psi$, and consequently the $\hat q_{n,m}^{i,j}$ in~\eqref{eq:xiim} satisfies
$
|\hat q_{n,m}^{i,j}| \le \hat \psi \triangleq A\psi  \int_{-\infty}^\infty |\Omega(t)|dt.
$
Based on this, we consider
\begin{equation*}
\begin{aligned}
&\mathbb{E}[|\xi_{i,m}|^2] \\
 \approx & \sum_{j=1}^J   |1-e^{-\jj \omega_{i,j}}|^2  \sum_{n=1}^{N-1} |\alpha_{i,j}|^2 \mathbb E[|\hat q_{n,m}^{i,j}|^2]   +  \sum_{j=1}^J |\alpha_{i,j}|^2\mathbb E[|\hat q_{N,m}^{i,j}|^2]\\
 \approx & \frac{4(N-1) \hat \psi^2}{3}   \sum_{j=1}^J |\alpha_{i,j}|^2  \sin^2 \left( \frac{\pi d}{\lambda} \sin(\theta_{i,j})\right) +  \frac{\hat \psi^2}{3} \sum_{j=1}^J \left| \alpha_{i,j}\right|^2,
\end{aligned}
\end{equation*}
where the  second line assumes that $\hat q_{n,m}^{i,j}$'s are zero-mean and independent of each other, and the last line assumes that $\hat q_{n,m}^{i,j}$'s are i.i.d. with $|\hat q_{n,m}^{i,j}| \sim \mathcal{U}{[0,\hat \psi]}$ and $\arg(\hat q_{n,m}^{i,j}) \sim \mathcal{U}{[-\pi,\pi]}$.

Then, we assume that the combined noise term $\eta_{i,m} = \xi_{i,m} +  v_{i,m}$ in~\eqref{eq:dist_time_mod} has $\eta_{i,m} \sim \mathcal{CN}(0,\sigma^2_{\eta,i})$, where $\sigma^2_{\eta,i} = \mathbb{E}[|\xi_{i,m}|^2] + \sigma_v^2$.
As a result, by the time-frequency transformation, one can show that the frequency-domain system model in~\eqref{eq:freq_mod} is simplified as
\begin{equation}\label{eq:freq_mod_approx}
	r_{i,p} = \check \bh_{i,p}^\Tsf \bz_p+ \check \eta_{i,p}, 
\end{equation}
where $\check \eta_{i,p} = \check \xi_{i,p} + \check v_{i,p}$ and $ \check \eta_{i,p} \sim \mathcal{CN}(0,\sigma^2_{\eta,i})$.
Note that if one consider the T$\Sigma\Delta$
scheme in Section~\ref{sec:sd_tr}, then
\[
\mathbb{E}[|\xi_{i,m}|^2] \approx \frac{4(N-1) \hat \psi^2}{3}   \sum_{j=1}^J |\alpha_{i,j}|^2  \sin^2 \left( \frac{\pi d}{\lambda} \sin(\theta_{i,j})\right),
\]
and the remaining derivation follows the same spirit  as before.
Next, we characterize the DP under the signal model in~\eqref{eq:freq_mod_approx}.
Each user detects the symbols by
\begin{equation}\label{eq:dec}
	\hat{s}_{i,p} = {\rm dec} \left( \frac{r_{i,p}}{\beta_i} \right),
\end{equation}
where $\hat{s}_{i,p}$ is the detected symbol of $s_{i,p}$ and ${\rm dec}(\cdot)$ is the decision function corresponding to the QAM constellation set $\setS$.
Given the data symbols $s_{i,p}$'s, the DP is defined as the probability of the correct detection of all the symbols, i.e.,
\begin{equation}\label{eq:dp}
	\begin{aligned}
		{\sf DP} &= \textstyle  {\rm Pr}\left( \hat{s}_{i,p} = s_{i,p}, \forall i,p \mid  \{s_{i,p}\}_{i,p} \right) \\
		&= \textstyle \prod_{i=1}^K \prod_{p=0}^{M_s-1} {\rm Pr}( \hat{s}_{i,p} = s_{i,p} \mid  s_{i,p} ) \\
		&= \textstyle \prod_{i=1}^K \prod_{p=0}^{M_s-1} \left( {\sf DP}^R_{i,p} \times {\sf DP}^I_{i,p} \right),
	\end{aligned}
\end{equation}
where
\begin{equation*}
	\begin{split}
		{\sf DP}^R_{i,p} &\triangleq {\rm Pr}( \Re(\hat{s}_{i,p}) = \Re(s_{i,p}) \mid s_{i,p} ), \\
		{\sf DP}^I_{i,p} &\triangleq {\rm Pr}( \Im(\hat{s}_{i,p}) = \Im(s_{i,p}) \mid s_{i,p} )
	\end{split}
\end{equation*}
represent the detection probabilities for the real and imaginary components of $s_{i,p}$, respectively.
Following the derivations in~\cite{shao2019framework,liu2021symbol}, it is shown that
\begin{equation} \label{eq:CDP}
	{\sf DP}^R_{i,p} \!=\!
	\begin{cases}
		\Phi \! \left( \!\frac{\sqrt{2}a_{i,p}^R}{\sigma_{\eta,i}}\! \right) \!-\! \Phi \!\left(\! \frac{\sqrt{2}c_{i,p}^R}{\sigma_{\eta,i}} \!\right),~ | \Re(s_{i,p}) | \!<\! 2D - 1,\\[2ex]
		\Phi \! \left(\! \frac{-\sqrt{2}c_{i,p}^R}{\sigma_{\eta,i}} \!\right), ~ \Re(s_{i,p}) \!=\!  2D - 1, \\[2ex]
		\Phi \! \left(\! \frac{\sqrt{2}a_{i,p}^R}{\sigma_{\eta,i}} \!\right), ~ \Re(s_{i,p}) \!=\!  -2D + 1,
	\end{cases}
\end{equation}
where $\Phi(x) = \frac{1}{\sqrt{2\pi}} \int_{-\infty}^x e^{-{y^2}/{2}}dy$ and
\begin{equation*}
	\begin{split}
		a_{i,p}^R &=  \beta_i + \beta_i\Re(s_{i,p})  - \Re(\check \bh_{i,p}^\Tsf \bz_p), \\
		c_{i,p}^R &= -\beta_i + \beta_i\Re(s_{i,p})  - \Re(\check \bh_{i,p}^\Tsf \bz_p).
	\end{split}
\end{equation*}
The result in \eqref{eq:CDP} also applies to ${\sf DP}^I_{i,p}$ by replacing ``$R$'' with ``$I$'' and ``$\Re$'' with ``$\Im$''.


We formulate an SLP formulation that maximizes the DP in \eqref{eq:dp} subject to the no-overloading condition in \eqref{eq:no_overloading}:
\begin{equation}\label{eq:SLP_origin}
	\begin{aligned}
		\max_{  \bm \beta, \bZ, \bX }
		& ~ \textstyle \prod_{i=1}^K \prod_{p=0}^{M_s-1}\left( {\sf DP}^R_{i,p} \times {\sf DP}^I_{i,p} \right) \\
		{\rm s.t.} ~
		&  ~ \bX = \bZ \bF_s^\Tsf,~ \bX \in \setX,~ \bm \beta \ge \bm 0,
	\end{aligned}
\end{equation}
where $\bm \beta = [\beta_1,\dots,\beta_K]^\Tsf$,
$\bZ = [\bz_0,\dots,\bz_{M_s-1}]$,
and $\bX = \bZ \bF_s^\Tsf$ is due to equation~\eqref{eq:bar_bx_n}.
A convenient way of handling the objective function in problem~\eqref{eq:SLP_origin} is to apply the $\log$ function over it, which results in
\begin{equation}\label{eq:SLP}
	\begin{aligned}
		\min_{ \bm \beta, \bZ, \bX }
		& ~
		F(\bm \beta, \bZ) \triangleq -
		\sum_{i=1}^{K} \sum_{p=0}^{M_s-1} \left(\log {\sf DP}^R_{i,p} + \log {\sf DP}^I_{i,p} \right) \\
		{\rm s.t.} ~
		& ~ \bX = \bZ \bF_s^\Tsf,~ \bX \in \setX,~ \bm \beta \ge \bm 0.
	\end{aligned}
\end{equation}

\noindent
Problem~\eqref{eq:SLP} is a large-scale convex problem.
It is large-scale because $M$, $M_s$ and $N$ are large under typical massive MIMO-OFDM system settings;
for example, when $(M,M_s,N)=(512,350,64)$, the dimensions of both $\bX$ and $\bZ$ will be over $20,000$.
This large-scale problem nature prevents us from applying general convex optimization tools, such as CVX~\cite{grant2008cvx}, to solve the problem, as the computational complexity will be very high.
Therefore, fast algorithms for problem~\eqref{eq:SLP} are desired, which motivates the last step as follows.

As the last step,
we custom design an alternating direction method of multipliers (ADMM) algorithm for problem~\eqref{eq:SLP}.
According to~\cite{boyd2011distributed}, ADMM is guaranteed to converge to an optimal solution of problem~\eqref{eq:SLP} under some mild conditions.
The development details are as follows.
The augmented Lagrangian of problem~\eqref{eq:SLP} is
\begin{equation*}
	\begin{aligned}
		L_{\rho}(\bm \beta,\bZ,\bX, \bm \Lambda) \!\!=\!\! \textstyle F(\bm \beta, \bZ)\!\! + \!\!  \langle \bX \!\!-\!\! \bZ \bF_s^\Tsf, \bm \Lambda \rangle \!\!+ \!\! \frac{\rho}{2}\| \bX \!\!-\!\! \bZ \bF_s^\Tsf \|_F^2,
	\end{aligned}
\end{equation*}
where
$\rho > 0$ is a penalty parameter and
$\bm \Lambda \in \Cbb^{N \times M}$ denotes the dual variables.
The ADMM iterations are given by
\begin{subequations}\label{eq:admm}
	\begin{align}
		\bX^{k+1} &= \arg\min\limits_{\bX \in \setX} L_{\rho}(\bm \beta^k,\bZ^k,\bX, \bm \Lambda^{k}), \label{eq:admm_pro1}\\
		(\bm \beta^{k+1} ,\bZ^{k+1}) &= \arg\min\limits_{\bm \beta \ge \bm 0} L_{\rho}(\bm \beta,\bZ,\bX^{k+1}, \bm \Lambda^{k}), \label{eq:admm_pro2}\\
		\boldsymbol{\Lambda}^{k+1} &= \boldsymbol{\Lambda}^{k} + \rho (\bX^{k+1} - \bZ^{k+1} \bF_s^\Tsf),\label{eq:admm_pro3}
	\end{align}
\end{subequations}
for $k=0,1,\dots$, where $(\bm \beta^0,\bZ^0,\bX^0, \bm \Lambda^0)$ is the initialization.
It remains to specify the updates in \eqref{eq:admm_pro1} and \eqref{eq:admm_pro2}.
The update in~\eqref{eq:admm_pro1} is essentially a projection step
\begin{align*}
	\bX^{k+1} &= \Pi_{\setX} \left( \bZ^{k} \bF_s^\Tsf -  \textstyle \frac{1}{\rho} \bm \Lambda^{k}\right),
\end{align*}
where the projection $\Pi_{\setX}(\bx) \triangleq \arg\min_{\by\in \setX}\| \bx -\by \|_2^2$ admits a closed-form expression, i.e., if $\hat \bX = \Pi_{\setX}(\bX)$, then we have
\[
\hat x_{n,m} = \min\{ |x_{n,m}|, \chi - \psi \} e^{\jj \cdot \arg(x_{n,m})}, \quad \forall n,m.
\]
The problem in~\eqref{eq:admm_pro2} is a smooth convex problem with non-negative constraints and we solve the problem by the accelerated proximal gradient (APG) method~\cite{beck2017first}.
For convenience, let us express problem~\eqref{eq:admm_pro2} as
\begin{equation}\label{eq:problem_p}
	\min_{\bm \beta \ge \bm 0} h(\bm \beta, \bZ),
\end{equation}
where $h(\bm \beta, \bZ) = L_{\rho}(\bm \beta,\bZ,\bX^{k+1}, \bm \Lambda^{k})$.
The APG updates for problem~\eqref{eq:problem_p} are
\begin{equation}\label{eq:apg}
	\begin{split}
        \bm \beta^{l+1} &= \max\{ \bm 0, \bm \beta_{\sf ex}^l - \gamma_l \nabla_{\bm \beta}h(\bm \beta_{\sf ex}^l, \bZ_{\sf ex}^l) \}, \\
        \bZ^{l+1} &= \bZ_{\sf ex}^l - \gamma_l \nabla_{\bZ}h(\bm \beta_{\sf ex}^l, \bZ_{\sf ex}^l), \\
	\end{split}
\end{equation}
for $l = 0,1,\dots$
Here, $\bm \beta_{\sf ex}^l$ and $\bZ_{\sf ex}^l$ are the extrapolated points given by
\begin{align*}
\bm \beta_{\sf ex}^l &=  \textstyle {\bm \beta}^{l} + \frac{\mu_{l-1}-1}{\mu_l} ( \bm \beta^{l} - \bm \beta^{l-1}),\\
\bZ_{\sf ex}^l &=  \textstyle {\bZ}^{l} + \frac{\mu_{l-1}-1}{\mu_l} ( \bZ^{l} - \bZ^{l-1}),
\end{align*}
where
$\bm \beta^{-1} = \bm \beta^{0}$, $\bZ^{-1} = \bZ^0$, $\mu_{-1} = 0$ and $\mu_{l} = \frac{1+\sqrt{1+4\mu_{l-1}^2}}{2}$;
$\gamma_l$ is the step size and is determined by the backtracking line search~\cite{beck2017first}.

We should mention the big-O computational complexity of the ADMM algorithm in \eqref{eq:admm}.
The complexity of performing the updates in \eqref{eq:admm_pro1} and \eqref{eq:admm_pro3} is $\mathcal{O}(NM \log_2 M)$, which mainly comes from computing
the inverse fast Fourier transform (IFFT).
For the update in \eqref{eq:admm_pro2}, performing one APG iteration in \eqref{eq:apg} requires $\mathcal{O}(KNM_s)$ floating point operations and $\mathcal{O}(KM_s)$ of computing $\Phi$.
Note that $\Phi$ does not have a closed-form expression; its computation can be done by the {\sf erfc} function in MATLAB and consumes more than one floating point operation.

\section{Simulation Results}
\label{sec:simulation}

In this section, we examine the performance of the developed $\Sigma \Delta$ precoding schemes by numerical simulations.

\subsection{Simulation and Algorithm Settings}

The default simulation settings are as follows.
The IDFT/DFT size is $M=512$.
The number of subcarriers is $M_s=300$.
The symbols $s_{i,p}$'s are uniformly drawn from the QAM constellation.
The adopted PA model is the modified Rapp model in \eqref{eq:rapp} with $A=16$, $\varphi=1.1$, $r_{\rm max}=0.1187$, $B=-345$, $C=0.17$ and $\zeta =4$~\cite{3gpp_pa}.
The inter-antenna spacing is $d = \lambda / 8$.
The numbers of channel paths and taps are $J = 4$ and $L = 20$, respectively.
For each channel path, the gain, angle and delay are generated by $\alpha_{i,j} \sim \frac{1}{\sqrt{J}} \mathcal{CN}(0,1)$,  $\theta_{i,j} \sim \mathcal{U}{[-35^\circ,35^\circ]}$ and $\tau_{i,j} \sim \mathcal{U}{[5T_s,15T_s]}$, respectively.
The receive low-pass filter $\Omega(t)$ in \eqref{eq:yim} is the root-raised cosine filter with period $T_s$ and roll-off factor $0.22$~\cite{mollen2016waveforms}.
All the continuous-time signals are represented by 7 times oversampled discrete signals, i.e., $7$ uniform sampling points for every interval $T_s$.
Unless specified, the results to be presented are averaged results over $1000$ independent channel trials.

\begin{table*}
	\centering
	\renewcommand{\arraystretch}{1.5}
	\caption{Summary of the tested precoding schemes.}\label{tab_schemes}
	\resizebox{\linewidth}{!}{
		\begin{tabular}{c|c|c|c|c|c}
			\hline
			name & description &  constraint for $\bX$ & antennas w/ distortion  &  w/ or w/o  $\Sigma \Delta$   & formulation, method\\
			\hline \hline
			$\Sigma \Delta$ ZF  &  ZF in Section~\ref{sec:precoding} w/ $\Sigma \Delta$ & \eqref{eq:no_overloading}  & all  & w/ $\Sigma \Delta$ & \multirow{6}{*}{\thead{\eqref{eq:zf} w/ specific $\Gamma$,\\ closed-form}}  \\ \cline{1-5}
			T$\Sigma \Delta$ ZF  & ZF in Section~\ref{sec:precoding} w/ T$\Sigma \Delta$  & \eqref{eq:no_overloading} & except the last antenna & w/ $\Sigma \Delta$ &  \\ \cline{1-5}
			ZF w/o $\Sigma \Delta$   & ZF in Section~\ref{sec:precoding} w/o $\Sigma \Delta$  & \eqref{eq:no_overloading} & all & w/o $\Sigma \Delta$ &  \\ \cline{1-5}
			ZF w/o distortion  & ZF in Section~\ref{sec:precoding} w/o distortion & \eqref{eq:no_overloading} & none  & w/o $\Sigma \Delta$ &  \\ \cline{1-5}
			ZF-TP  & total power constrained ZF & \eqref{eq:ZF_TPC} & except the last antenna  & w/o $\Sigma \Delta$ & \\ \cline{1-5}
			ZF-BO  & ZF w/ power back-off & \eqref{eq:const_1db}  & except the last antenna  & w/o $\Sigma \Delta$ &  \\ \hline
			\hline
			$\Sigma \Delta$ SLP  & SLP in Section~\ref{sec:precoding}  w/ $\Sigma \Delta$ & \eqref{eq:no_overloading} & all  & w/ $\Sigma \Delta$ & \multirow{4}{*}{\thead{\eqref{eq:SLP} w/ specific $\setX$,\\ ADMM in~\eqref{eq:admm}}} \\ \cline{1-5}
			T$\Sigma \Delta$ SLP   & SLP in Section~\ref{sec:precoding} w/ T$\Sigma \Delta$  & \eqref{eq:no_overloading}  & except the last antenna  & w/ $\Sigma \Delta$ &  \\ \cline{1-5}
			SLP w/o distortion & SLP in Section~\ref{sec:precoding} w/o distortion & \eqref{eq:no_overloading} & none  & w/o  $\Sigma \Delta$ &  \\ \cline{1-5}
			SLP-BO  & SLP w/ power back-off &  \eqref{eq:const_1db}  & except the last antenna & w/o $\Sigma \Delta$ &   \\ \hline
		\end{tabular}
	}
\end{table*}

For clarity, we summarize all the tested precoding schemes in Table~\ref{tab_schemes}.
Three benchmark precoding schemes are considered, namely, ZF with total power constraints (ZF-TP), ZF with power back-off (ZF-BO) and SLP with power back-off (SLP-BO).
The ZF-TP scheme takes the ZF form in~\eqref{eq:zf} and the normalization factor $\Gamma$ is chosen to ensure the total power constraint
\begin{equation}\label{eq:ZF_TPC}
	\Exp[\|\bX\|_F^2] = NMr_{\rm max}^2.
\end{equation}
The ZF-BO scheme also takes the form~\eqref{eq:zf}, but the $\Gamma$ is chosen to satisfy the constraint in \eqref{eq:1db}, such that the PAs will operate in the linear regions.
It follows from~\eqref{eq:OFDM_sig} that the constraint in \eqref{eq:1db} is satisfied if and only if
\begin{equation}\label{eq:const_1db}
	\bX \in \setX_{\rm 1dB} = \{ \bX \in \Cbb^{N \times M} ~\vert~  \|\bX\|_{\rm max} \le r_{\rm 1dB} \},
\end{equation}
and thus the $\Gamma$ is given by
\[
\Gamma = \frac{\| [\check\bH_0^\dagger\bs_0,\dots,\check\bH_{M_s-1}^\dagger\bs_{M_s-1}] \bF_s^\Tsf \|_{\rm max} }{r_{\rm 1dB}}.
\]
Following the same spirit, the SLP-BO scheme is formulated as problem~\eqref{eq:SLP} with $\setX = \setX_{\rm 1dB}$, which is solved by the ADMM algorithm in~\eqref{eq:admm}.
In addition, we consider the ZF scheme and the SLP scheme without PA distortions (and  without $\Sigma \Delta$ modulation); the ZF scheme with PA distortions but without $\Sigma \Delta$ modulation is also included to show the impact of PA distortion.

The settings of the ADMM algorithm in \eqref{eq:admm} are as follows.
The initialization for the primal variables $(\bm \beta^0, \bZ^0, \bX^0)$ is chosen as the ZF scheme in Section~\ref{sec:precoding}.
The initialization for the dual variables is $\bm \Lambda^0=\bzero$.
The penalty parameter is fixed as $\rho=500$.
The algorithm is terminated when $|F(\bm \beta^{k+1}, \bm \bZ^{k+1})-F(\bm \beta^{k}, \bm \bZ^{k})| \le 10^{-3}F(\bm \beta^{k}, \bm \bZ^{k})$ and $\|\bX^{k+1} - \bZ^{k+1} \bF_s^\Tsf\|_F^2 \le 10^{-3}$, or when the number of ADMM iterations exceeds $30$.
The APG method in \eqref{eq:apg} for subproblem~\eqref{eq:admm_pro2} is initialized with the previous iterate $(\bm \beta^k, \bZ^k)$ and stops when $\|\bm \beta^{l+1}-\bm \beta^l\|_2^2 +  \|\bZ^{l+1}-\bZ^l\|_F^2\le 10^{-6}$ or when the number of APG iterations exceeds $50$.

\begin{figure}[t]
	\centering
	\subfigure[ZF w/o distortion]{
		\includegraphics[width=0.4\linewidth]{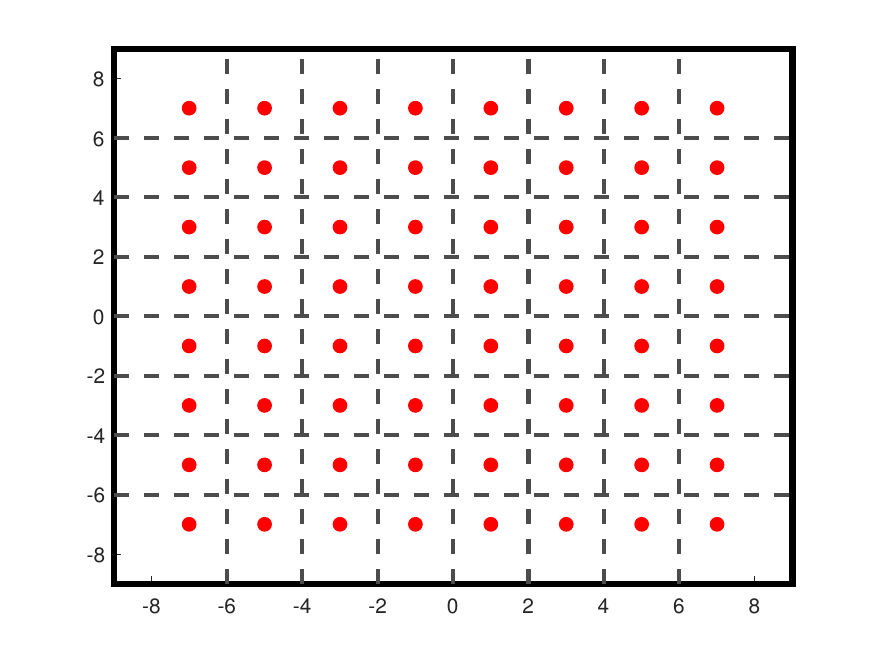}}
	\subfigure[ZF w/o $\Sigma \Delta$]{
		\includegraphics[width=0.4\linewidth]{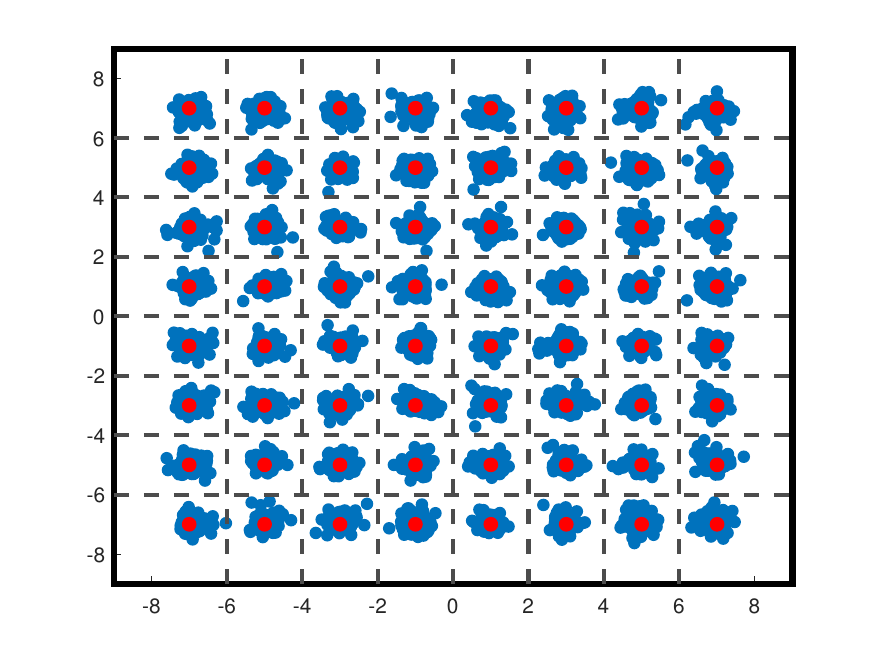}}\\
	\subfigure[$\Sigma \Delta$ ZF]{
		\includegraphics[width=0.4\linewidth]{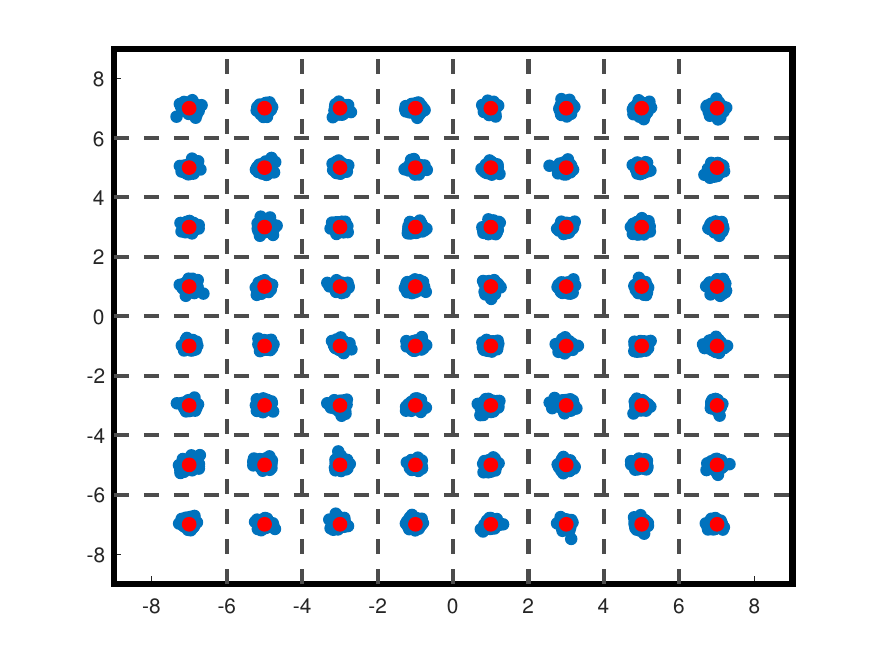}}
	\subfigure[T$\Sigma \Delta$ ZF]{
		\includegraphics[width=0.4\linewidth]{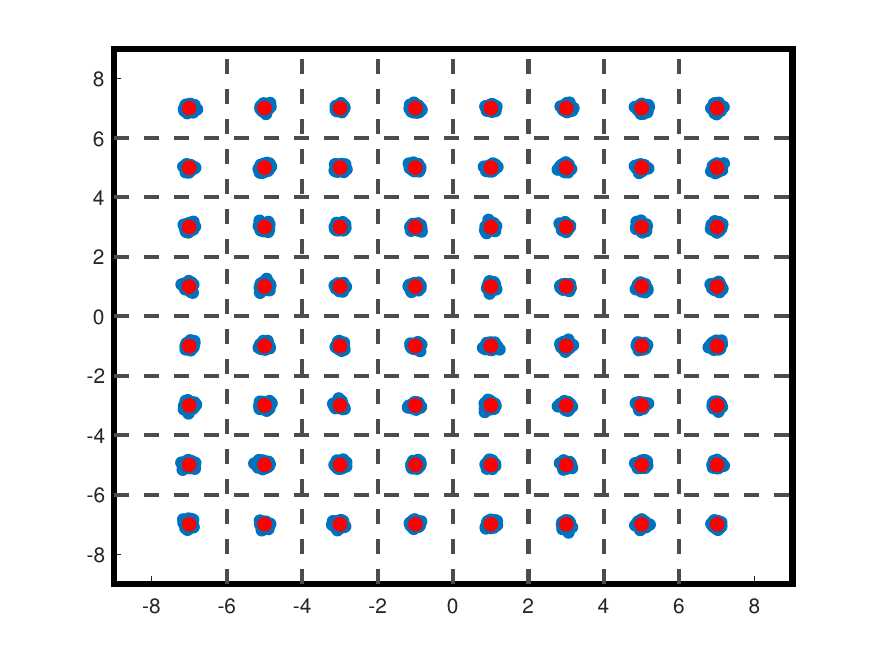}}
	\caption{IQ scatter plots. ${(N,K) = (64,10)}$.}
	\label{fig_scatter_zf_64_10}
\end{figure}

\begin{figure}[t]
	\centering
	\subfigure[ZF w/o distortion]{
		\includegraphics[width=0.4\linewidth]{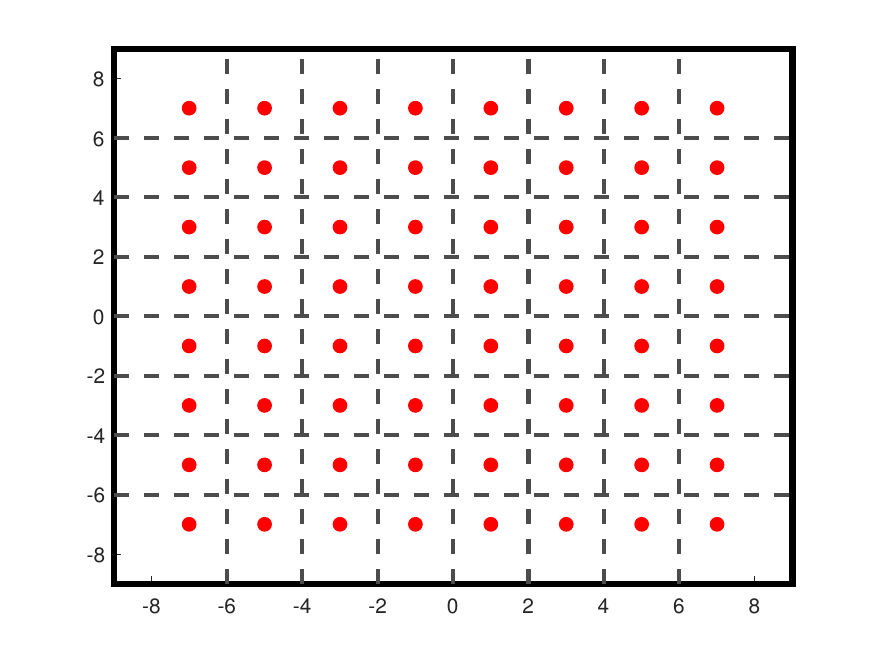}}
	\subfigure[ZF w/o $\Sigma \Delta$]{
		\includegraphics[width=0.4\linewidth]{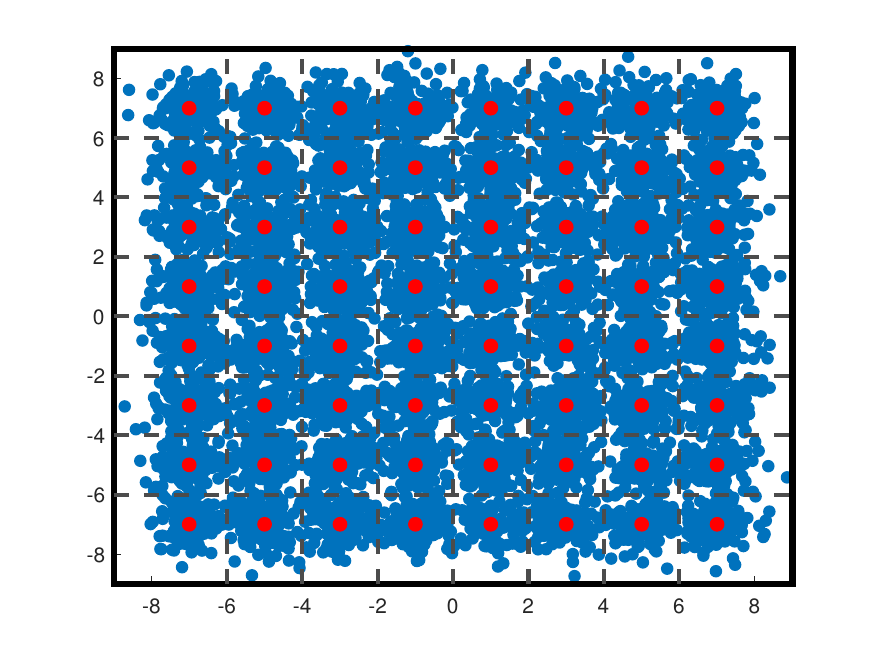}}\\
	\subfigure[$\Sigma \Delta$ ZF]{
		\includegraphics[width=0.4\linewidth]{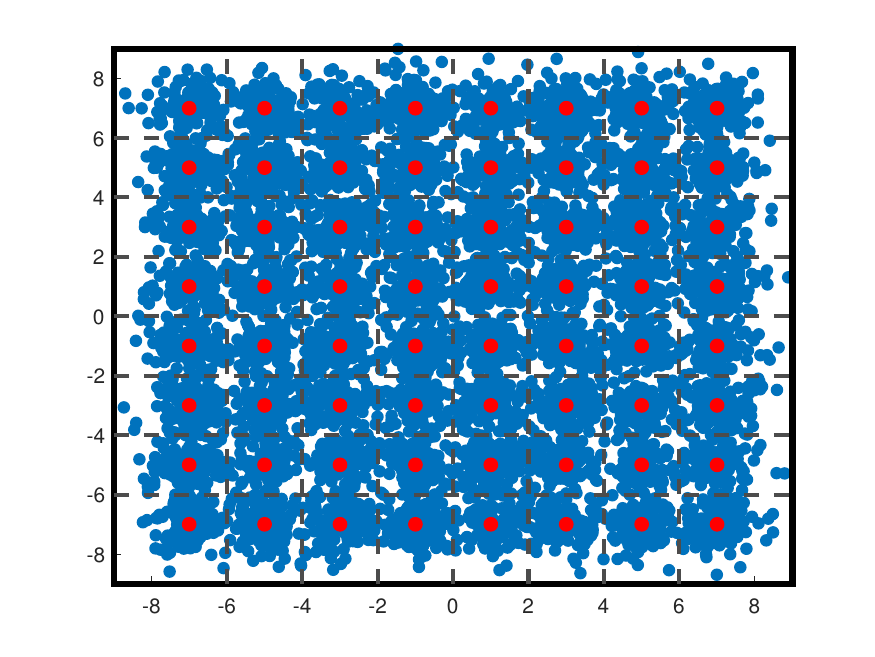}}
	\subfigure[T$\Sigma \Delta$ ZF]{
		\includegraphics[width=0.4\linewidth]{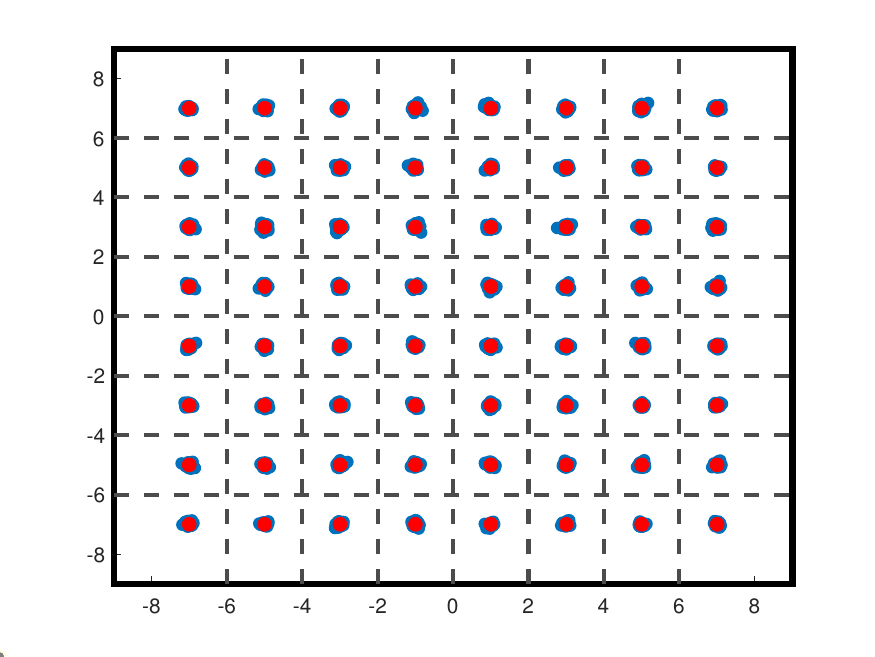}}
	\caption{IQ scatter plots. ${(N,K) = (56,10)}$.}
	\label{fig_scatter_zf_56_10}
\end{figure}

\subsection{IQ Scatter Plots}

We begin with presenting the in-phase quadrature-phase (IQ) scatter plots.
Fig.~\ref{fig_scatter_zf_64_10} shows the results of the ZF schemes for $(N,K) = (64,10)$.
The red dots represent the constellation points and the blue dots represent the background noise-free received signals at the first subcarrier, i.e., $r_{i,p}/\beta_i$ in~\eqref{eq:dec} with $p=1$ and without background noise.
The dashed lines are the decision boundaries.
We fix one channel and overlay the received signals for $1000$ OFDM blocks.
Comparing ZF w/o distortion and ZF w/o $\Sigma \Delta$, we see that the PA distortions can have a great impact on the received signals.
Comparing ZF w/o $\Sigma \Delta$ and $\Sigma \Delta$ ZF, we see that the spatial $\Sigma \Delta$ approach can help mitigate the PA distortion effect.
Comparing $\Sigma \Delta$ ZF and T$\Sigma \Delta$ ZF, we see that the mitigation effect of the spatial $\Sigma \Delta$ approach can be enhanced by employing a linear PA at the last antenna.

In Fig.~\ref{fig_scatter_zf_56_10}, we reduce the number of transmit antennas from $N=64$ to $N=56$ and show the IQ scatter plots.
In this case, T$\Sigma \Delta$ ZF still works well, but $\Sigma \Delta$ ZF fails to work.
This indicates that when the number of transmit antennas is relatively small, the PA distortion at the last antenna can have a notable effect on the received symbols.



\begin{figure*}
\centering
\subfigure[BERs of the ZF schemes. $(N,K) = (16,4)$.]{
\includegraphics[width=0.45\textwidth]{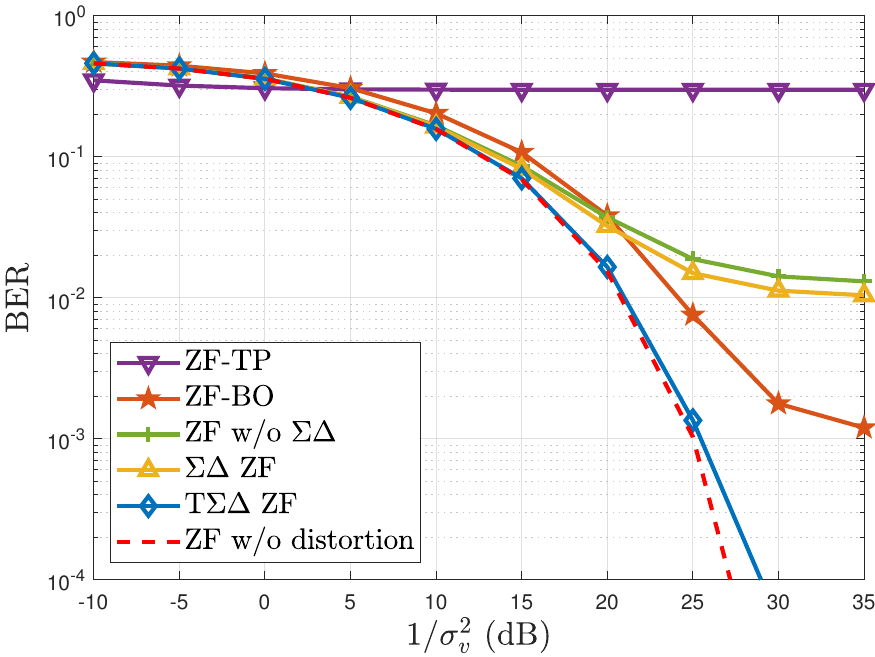}\label{fig_ber_zf_16_4}}
\subfigure[BERs of the SLP schemes. $(N,K) = (16,4)$.]{
\includegraphics[width=0.45\textwidth]{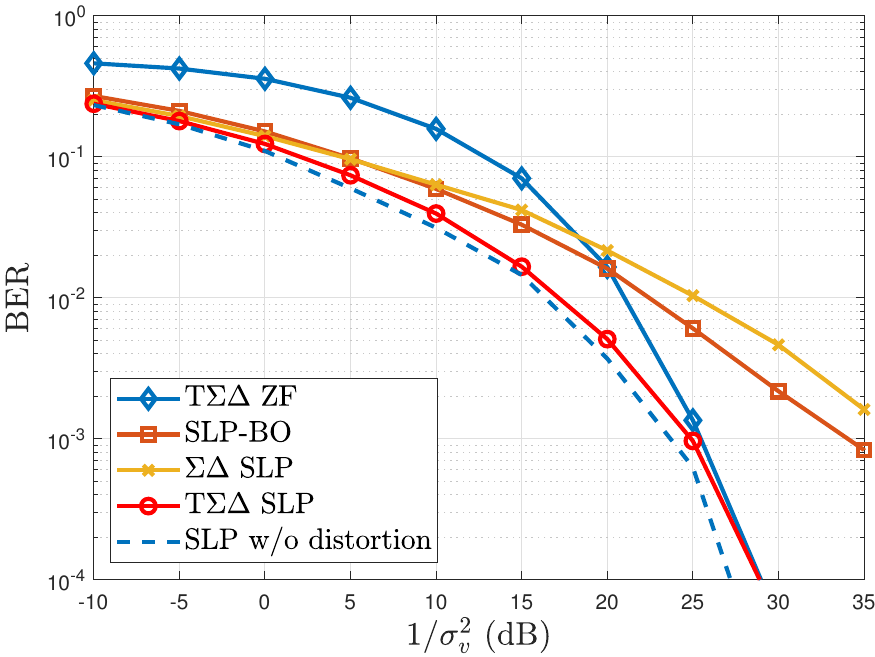}\label{fig_ber_slp_16_4}}
\subfigure[BERs of the ZF schemes. $(N,K) = (32,6)$.]{
\includegraphics[width=0.45\textwidth]{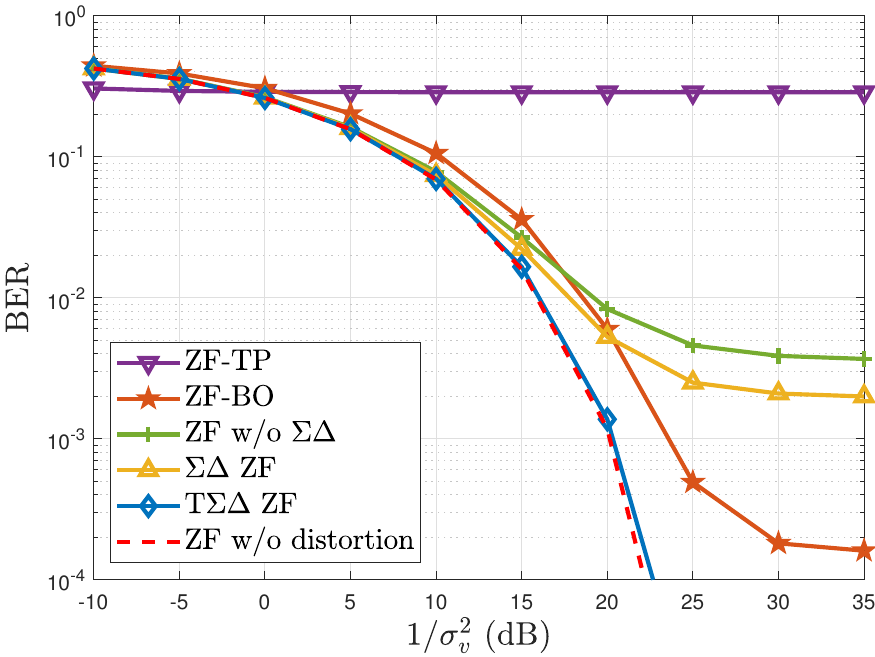}\label{fig_ber_zf_32_6}}
\subfigure[BERs of the SLP schemes. $(N,K) = (32,6)$.]{
\includegraphics[width=0.45\textwidth]{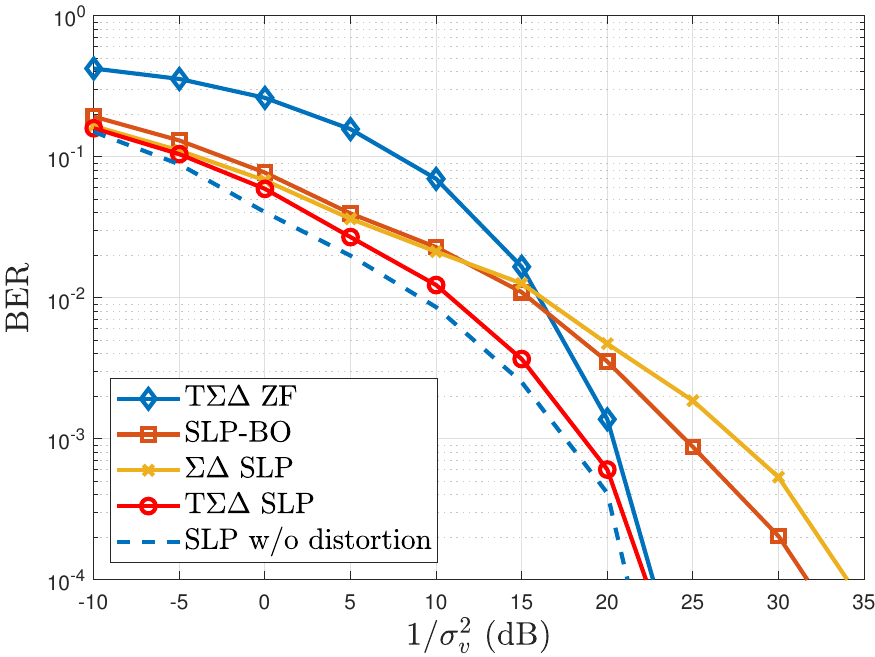}\label{fig_ber_slp_32_6}}
\subfigure[BERs of the ZF schemes. $(N,K) = (64,8)$.]{
\includegraphics[width=0.45\textwidth]{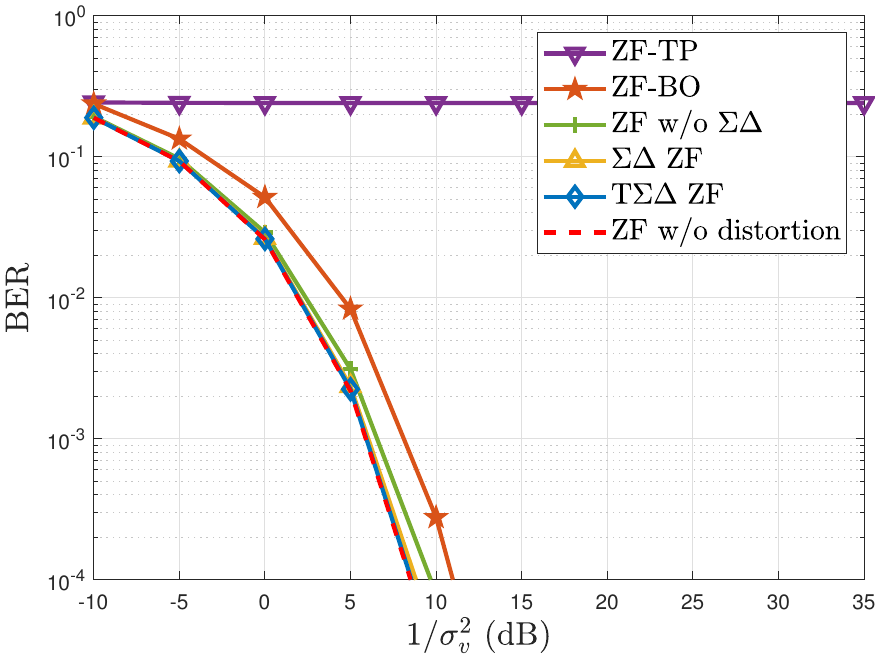}\label{fig_ber_zf_64_8}}
\subfigure[BERs of the SLP schemes. $(N,K) = (64,8)$.]{
\includegraphics[width=0.45\textwidth]{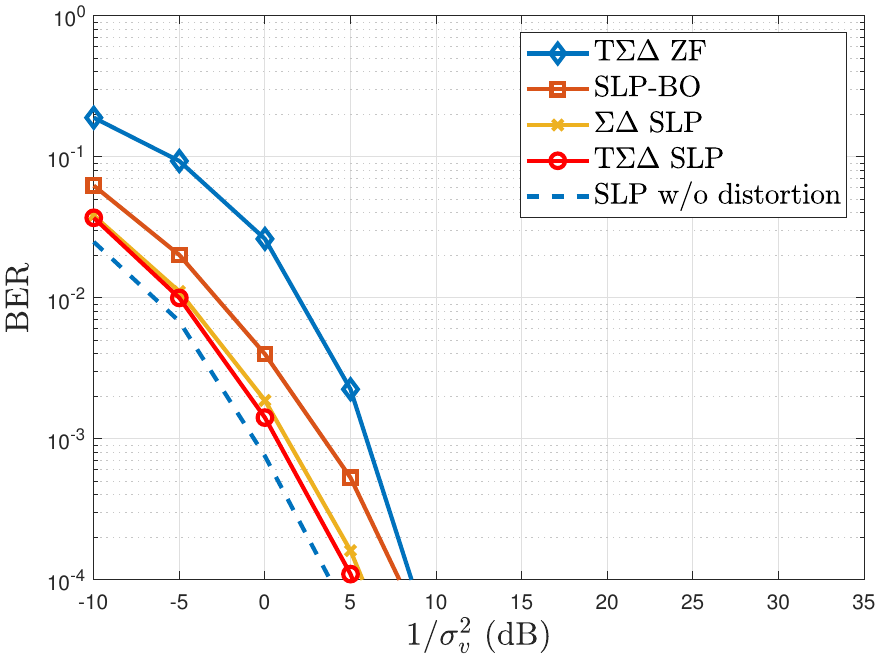}\label{fig_ber_slp_64_8}}
\caption{BERs under different system settings.}
\label{fig_ber}
\end{figure*}

%
%
%
%

\begin{figure*}
	\centering
	\subfigure[256-QAM.]{
		\includegraphics[width=0.45\linewidth]{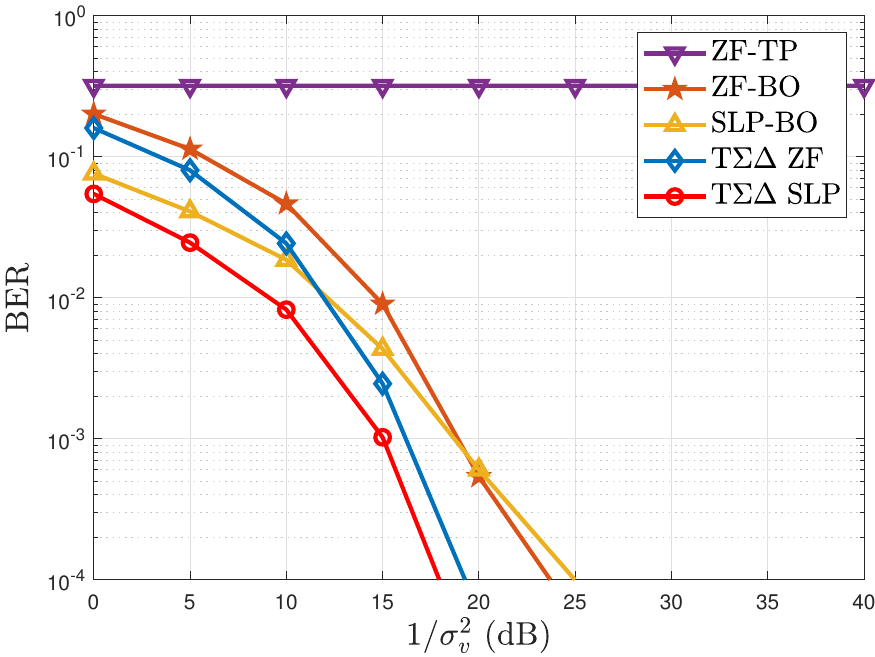}}
	\subfigure[1024-QAM.]{
		\includegraphics[width=0.45\linewidth]{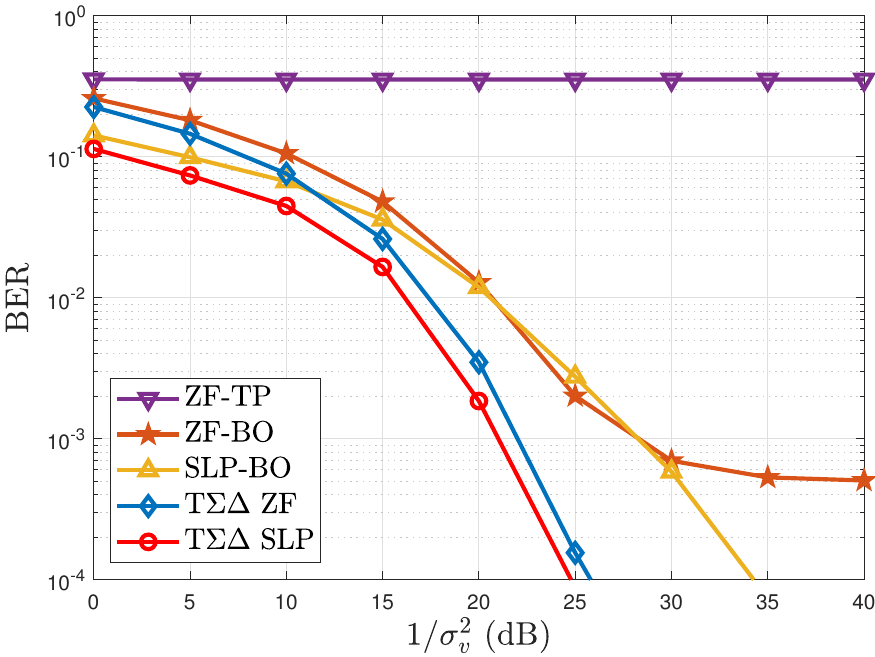}}
	\caption{BERs for high-order QAM. ${(N,K) = (56,8)}$.}
	\label{fig_ber_high_QAM}
\end{figure*}

\begin{figure*}
	\centering
	\subfigure[$\varphi = 1.0$]{
		\includegraphics[width=0.45\linewidth]{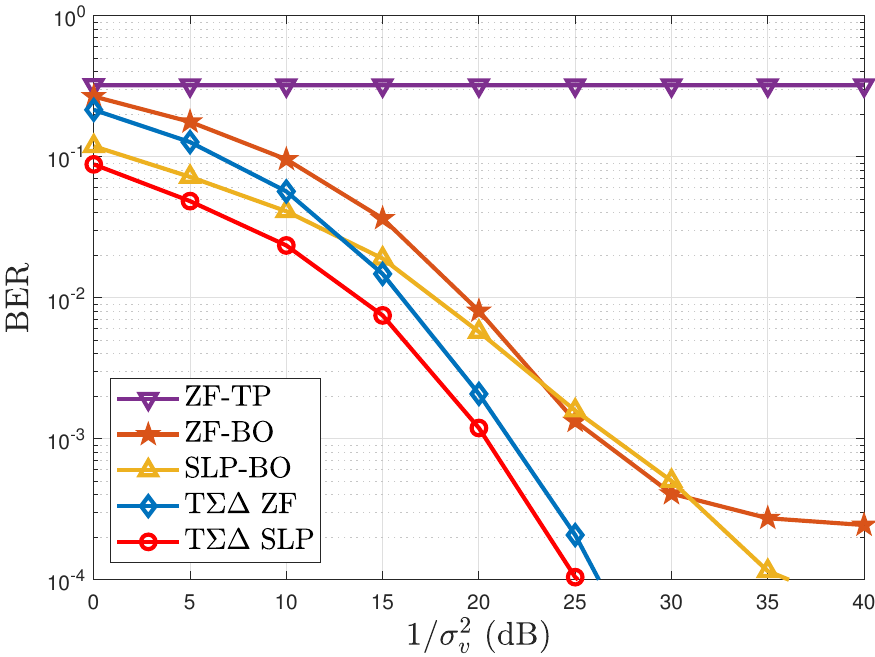}}
	\subfigure[$\varphi = 1.5$]{
		\includegraphics[width=0.45\linewidth]{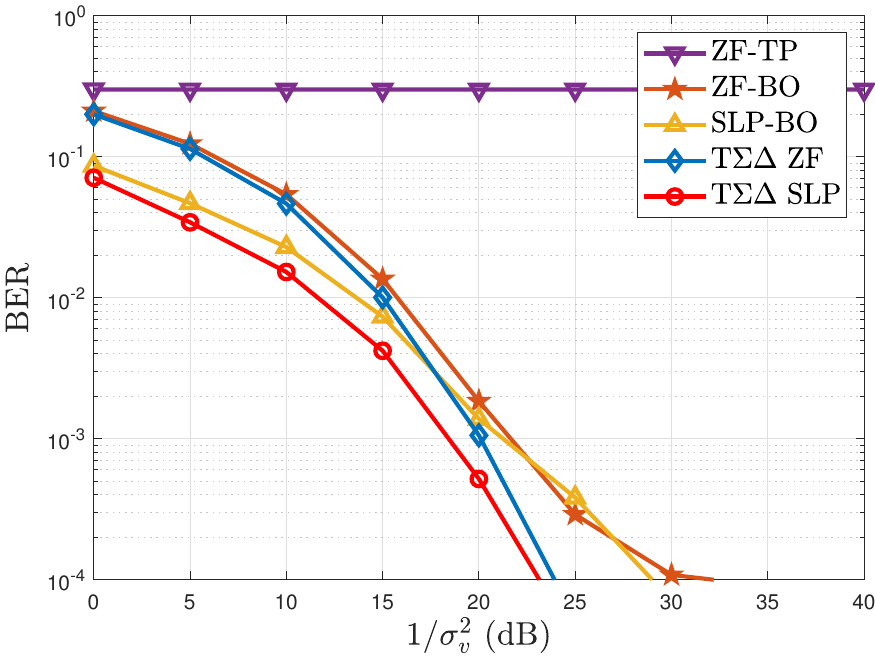}}
	\subfigure[$\varphi = 2.0$]{
		\includegraphics[width=0.45\linewidth]{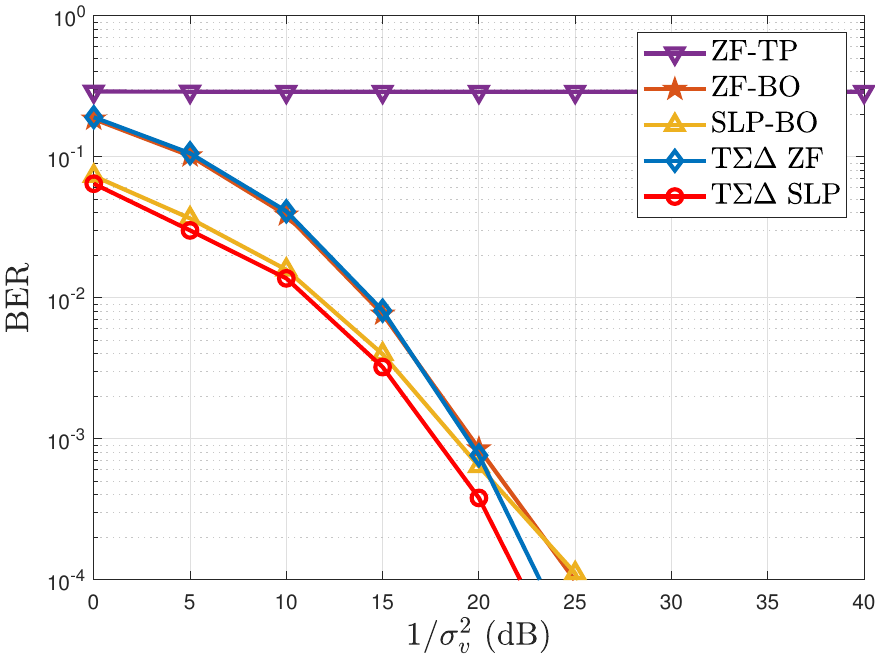}}
	\caption{BERs for different ${\varphi}$. ${(N,K) = (32,5)}$, 256-QAM.}
	\label{fig_ber_sspa}
\end{figure*}

\begin{figure}
	\centering
	\subfigure[64-QAM.]{
		\includegraphics[width=0.45\linewidth]{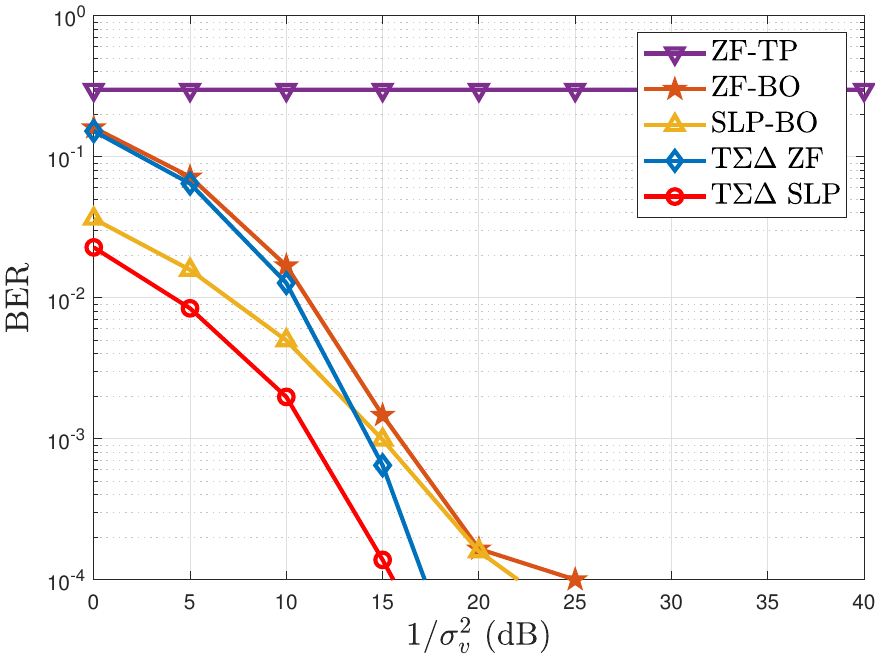}}
	\subfigure[256-QAM.]{
		\includegraphics[width=0.45\linewidth]{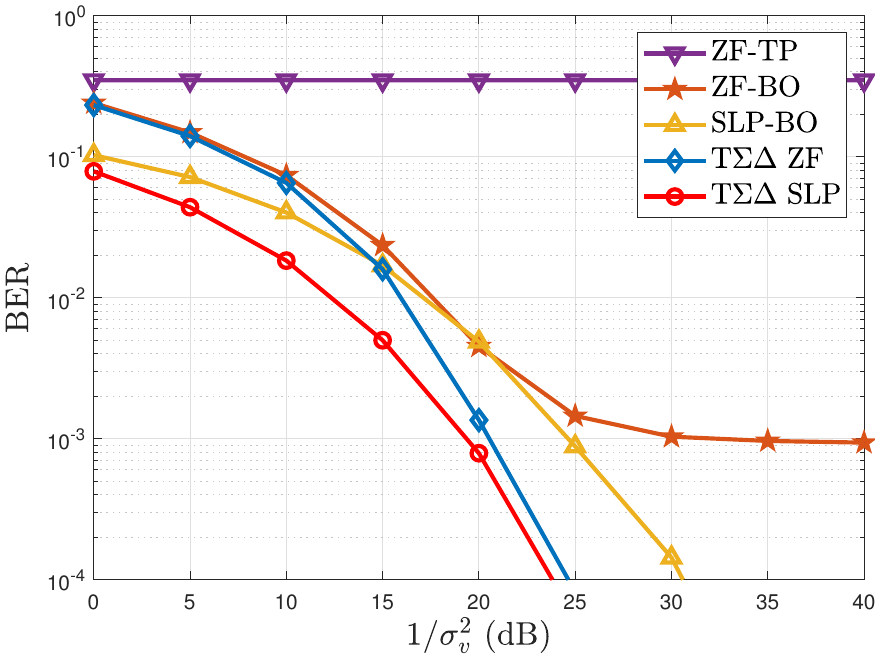}}
	\caption{BERs under the TWTA model. ${(N,K) = (64,10)}$.}
	\label{fig_ber_twta}
\end{figure}

\subsection{Bit Error Rates}

Now we turn to the bit error rate (BER) results.
Fig.~\ref{fig_ber} shows the BERs of the various precoding schemes under different problem sizes $(N,K)$.
Let us first focus on the case of the small problem size $(N,K)=(16,4)$.
For the ZF schemes as shown in Fig.~\ref{fig_ber_zf_16_4},
we observe that T$\Sigma\Delta$ ZF offers the most promising performance;
specifically, it can achieve similar BER performance as ZF w/o distortion and meanwhile outperform ZF-BO, ZF w/o $\Sigma \Delta$ and $\Sigma \Delta$ ZF.
However, $\Sigma \Delta$ ZF suffers from the error floor effect, mainly due to the PA distortion at the last antenna.
For the SLP schemes as shown in Fig.~\ref{fig_ber_slp_16_4}, we observe similar performance behaviors as those in Fig.~\ref{fig_ber_zf_16_4}.
Moreover, T$\Sigma \Delta$ SLP exhibits better performance than T$\Sigma \Delta$ ZF.
But it should be noted that compared with ZF, SLP demands higher computational costs for calling the ADMM algorithm.

In Fig.~\ref{fig_ber_zf_32_6}--Fig.~\ref{fig_ber_slp_64_8}, we show the BER performances for larger problem sizes, namely $(N,K)=(32,6)$ and $(N,K)=(64,8)$.
We see from Fig.~\ref{fig_ber_zf_32_6} and Fig.~\ref{fig_ber_zf_64_8} that the performance gap between ZF w/o distortion and ZF w/o $\Sigma \Delta$ tends to vanish as the problem size increases.
This suggests that the PA distortions have less influence on the BER performance.
We empirically find that as the problem size increases, the ZF w/o $\Sigma \Delta$, $\Sigma \Delta$ ZF and T$\Sigma \Delta$ ZF  schemes
tend to have larger QAM constellation scaling factors $\beta_i$'s, which means that the constellation points are more separated and intuitively it is more unlikely for the PA distortions to cause additional bit errors.
Also, from Fig.~\ref{fig_ber_slp_64_8}, we see that the performance gain of T$\Sigma \Delta$ SLP over T$\Sigma \Delta$ ZF can be large for large problem sizes.


Next, we consider  higher-order QAM constellations.
Fig.~\ref{fig_ber_high_QAM} shows the BERs for 256-QAM and 1024-QAM constellations under $(N,K) = (56,8)$.
It is seen that both T$\Sigma \Delta$ ZF and T$\Sigma \Delta$ SLP yield good performances, especially when $1/\sigma_v^2$ is large.
Also, compared to the benchmark schemes, T$\Sigma \Delta$ ZF and T$\Sigma \Delta$ SLP tend to be more advantageous for larger QAM sizes, as intuitively higher-order QAM constellations are more vulnerable to distortions and noise.



\subsection{Performance under Other PA Model Parameters or PA Models}

Our developed $\Sigma \Delta$ precoding schemes can work for a variety of PA models.
We first evaluate the performance
under other parameters of the modified Rapp model in \eqref{eq:rapp}.
In particular, we fix $A=16$, $r_{\rm max}=0.1187$ and $B=0$, which corresponds to the case without the phase distortion ($g_p(r)=0$), and vary the parameter $\varphi$.
Note that in such case,  larger $\varphi$ means lower PA nonlinearities
and the case with $\varphi \to \infty$ corresponds to the ideal PA model in~\eqref{eq:ideal_pa}.
Fig.~\ref{fig_ber_sspa} presents the BERs for different $\varphi$.
We consider $(N,K) = (32,5)$ and 256-QAM constellation.
It is observed that T$\Sigma \Delta$ SLP (respectively T$\Sigma \Delta$ ZF) outperforms SLP-BO (respectively ZF-BO), and the performance gap reduces with increasing $\varphi$, or with weaker PA nonlinearities.


Finally, we test the precoding schemes under other PA models.
Specifically, we consider the TWTA model with the following AM-AM and AM-PM conversions
\begin{equation*}
g_a(r) = \frac{Ar}{1+\frac{1}{4}(r/r_{\rm max})^2}, ~
g_p(r) = \frac{\pi}{12} \frac{(r/r_{\rm max})^2}{1+\frac{1}{4}(r/r_{\rm max})^2} ({\rm rad}).
\end{equation*}
Fig.~\ref{fig_ber_twta} shows the BERs for 64-QAM and 256-QAM constellations under the TWTA model with $A=16$ and $r_{\rm max}=0.1187$.
We set $(N,K) = (64,10)$.
It is observed that the T$\Sigma \Delta$ ZF and T$\Sigma \Delta$ SLP schemes can still provide reasonably good performances.

\section{Conclusion}

To conclude, we proposed a spatial $\Sigma \Delta$ modulation concept for
PA nonlinear distortion mitigation in the massive MIMO downlink scenario.
While $\Sigma \Delta$ modulation has been used to shape and mitigate quantization noise in temporal DACs/ADCs for decades, its adaptation to mitigate PA distortion in space is new to the best of our knowledge.
Our presented approach can effectively mitigate the PA distortion effects at the user side, under the assumption of a ULA at the BS operating in a low-pass angular sector and using a small inter-antenna spacing.
We
applied the spatial $\Sigma \Delta$ approach to the multi-user massive MIMO-OFDM downlink scenario
and custom designed a ZF scheme and an SLP scheme for the precoding design.
Simulation results revealed that the developed $\Sigma \Delta$ precoding schemes can provide promising performance in PA distortion effect suppression.

\section*{Appendix: Proof of Fact~\ref{Fac:no_overloading}}



We prove Fact~\ref{Fac:no_overloading} by induction.
When $n=1$, it is shown that
\begin{align}
	|b_1(t)| &= |{x}_1(t)|  \le \chi - \psi \le \chi, \label{eq:app1_1} \\
	|q_1(t)| &= |G(b_1(t))/A-b_1(t)| \le \psi, \label{eq:app1_2}
\end{align}
where \eqref{eq:app1_2} is due to \eqref{eq:app1_1} and the definition of $\psi$ in~\eqref{eq:C}.
When $n \ge 2$, suppose that $|b_{n-1}(t)| \le \chi$ and $|q_{n-1}(t)| \le \psi$.
Then it holds that
\begin{equation}\label{eq:app1_4}
	\begin{aligned}
		|b_{n}(t)| &= |{x}_n(t) - q_{n-1}(t)| \\
		&\le |{x}_n(t)| + |q_{n-1}(t)| \\
		&\le \chi.
	\end{aligned}
\end{equation}
By \eqref{eq:app1_4} and \eqref{eq:C}, we have
\[
|q_n(t)| = |G(b_n(t))/A- b_n(t)| \le \psi.
\]
Therefore, $|b_{n}(t)| \le  \chi$ and $|q_{n}(t)| \le  \psi$ hold for all $n$.

\bibliographystyle{IEEEtran}

\end{document}